\documentclass[
 reprint,
 amsmath,amssymb,
 aps,
 nofootinbib,
 superscriptaddress
]{revtex4-2}
\usepackage{graphicx}% Include figure files
\usepackage{bm}% bold math
\usepackage{hyperref}% add hypertext capabilities
\usepackage{booktabs} % nicer tables
\usepackage{physics}
\usepackage{xcolor}
\usepackage{multirow}
\usepackage{leftindex}
\usepackage{subcaption}
\usepackage{ulem} %For \sout command for editing purposes
\usepackage{float}

\newcommand{\R}[0]{\mathcal{R}}

\newcommand{\picwidth}{0.9\columnwidth}
\newcommand{\picgap}{\hspace{0.8cm}}
\newcommand{\m}{m_{\text{P}}}
\newcommand{\ra}{\mathrm{r}}
\renewcommand{\[}{\begin{equation}}
\renewcommand{\]}{\end{equation}}

\begin{document}

\title{DBI Softens Stiff-Phase Gravitational-Wave Spectra}

\author{Samuel S\'anchez L\'opez}
\email{samuel.lopez@iap.fr}
\affiliation{Institut d’Astrophysique de Paris, 98 bis boulevard Arago, F-75014 Paris, France}
\author{Eemeli Tomberg}
\email{eemeli.tomberg@uclouvain.be}
\affiliation{Cosmology, Universe and Relativity at Louvain (CURL), Institute of Mathematics and Physics, \\ University of Louvain, 2 Chemin du Cyclotron, 1348 Louvain-la-Neuve, Belgium
}
\author{Juan C. Bueno S\'anchez}
\email{juancarlos.bueno@upm.es}
\affiliation{Área de Física Aplicada, E.T.S. de Ingeniería y Diseño Industrial, Universidad Politécnica de
Madrid, 28012 Madrid, Spain.}

\date{\today}

\begin{abstract}
We investigate the effect of non-canonical kinetic terms on the present-day density spectrum of primordial gravitational waves (GWs), by considering the dynamics of a scalar field governed by the Dirac--Born--Infeld (DBI) action and a runaway potential. The field, responsible for a period of inflation, subsequently enters a kinetically dominated regime, where the DBI structure reduces the equation-of-state parameter relative to the canonical kination result $w=1$, thereby softening the blue tilt of the GW spectrum. Simultaneously, the energy density of the field redshifts at a slower rate, delaying radiation domination and prolonging the stiff phase. The interplay between these two effects can lead to a potentially observable GW signal that would otherwise remain beyond the reach of future experiments in the canonical case. We test our predictions by numerically solving the background and perturbations, from inflation until radiation domination, and find viable parameter space compatible with cosmic microwave background and Big Bang Nucleosynthesis constraints.
\end{abstract}

\maketitle

\section{Introduction}
Inflation~\cite{Guth:1980zm,Linde:1981mu,Albrecht:1982wi,Starobinsky:1980te} provides the initial conditions of the hot Big Bang via the production of primordial scalar perturbations, together with a stochastic background of gravitational waves (GWs). The former have been constrained to high precision through measurements of the temperature anisotropies in the cosmic microwave background (CMB)~\cite{Planck:2018vyg}, favouring a nearly scale-invariant power spectrum with a small red tilt. At these scales, the landscape of inflationary models has been significantly restricted to flat and concave potentials, within the simplest single-field slow-roll scenarios. However, measurements of scalar perturbations alone cannot determine the energy scale of inflation or uniquely identify the dynamics of the field driving it. In this context, the detection of primordial GWs, often regarded as a smoking gun for inflation~\cite{Bahr-Kalus:2022prj}, would provide an independent test of the theory, probing physics out of reach from scalar perturbations. 

Importantly, the present-day density spectrum of primordial GWs, $\Omega_{\rm GW}^0(f)$, not only carries information about the mechanism that produced them, but also about the expansion history of the post-inflationary Universe~\cite{Boyle:2005se,Tashiro:2003qp}. Indeed, tensor modes freeze after leaving the inflationary horizon, preserving the amplitude and tilt of the primordial tensor power spectrum until horizon re-entry, after which their energy density redshifts as radiation. The subsequent sub-horizon evolution depends on the background equation-of-state (EoS) parameter, meaning that $\Omega_{\rm GW}^0(f)$ can exhibit different spectral shapes, depending on the energy content of the Universe.  This is of particular relevance since the expansion history before Big-Bang-Nucleosynthesis (BBN) is largely unconstrained~\cite{Allahverdi:2020bys}, making GWs an extremely useful probe~\cite{Figueroa:2019paj}. 

In standard inflationary scenarios, a nearly scale-invariant tensor power spectrum leads to an approximately flat $\Omega_{\rm GW}^0(f)$ for modes re-entering the horizon during the radiation dominated era~\cite{Sahni:1990tx,Watanabe:2006qe}. However, this picture can significantly change if the post-inflationary Universe undergoes a stiff era~\cite{Figueroa:2019paj} with EoS parameter $1/3 < w \leq 1$, producing an enhancement in the density spectrum of GWs. During such a phase, the energy density of radiation redshifts as $\rho\propto a^{-4}$, while that of the dominant stiff component follows $\rho\propto a^{-3(1+w)}$, assuming a roughly constant $w$. Since $w>1/3$, radiation eventually catches up and becomes dominant, giving way to the hot Big Bang. 

In the canonical case, kinetic domination leads to a phase with $w=1$, a regime called kination~\cite{Spokoiny:1993kt,Joyce:1996cp,Joyce:1997fc}. Tensor modes re-entering the horizon during this period contribute to a blue-tilted GW spectrum with slope equal to unity~\cite{Giovannini:1998bp,Giovannini:1999bh,Sahni:2001qp}. Such an enhancement could improve the detectability prospects by future GW experiments. However, an excessively prolonged kination phase may lead to a GW energy density incompatible with the bounds on additional radiation during BBN~\cite{Chen:2024roo,Park:2026kmj} (see also Ref.~\cite{Eroncel:2025bcb}), so a delicate balance must be met for a signal to be potentially observable.

Post-inflationary stiff phases can arise in a variety of cosmological scenarios. One example is non-oscillatory (NO) models of inflation~\cite{Felder:1999pv}, where the potential has no minimum, becoming steep after inflation ends. Since reheating channels through inflaton oscillations are not available, some alternative mechanism produces the radiation bath. Such phases are also prominent in quintessential inflation~\cite{Peebles:1998qn} (see Refs.~\cite{Dimopoulos:2001ix,Dimopoulos:2020pas,Dimopoulos:2022tvn,Dimopoulos:2022rdp,Dimopoulos:2017zvq,Sami:2004xk} for representative examples). However, stiff phases can also take place in string compactifications~\cite{Apers:2022cyl,Conlon:2008cj,Conlon:2022pnx,Apers:2024ffe}, rotating axion fields~\cite{Co:2019jts,Harigaya:2023mhl,Harigaya:2023pmw,Duval:2024jsg,Chung:2024ctx,Co:2021lkc}, scalar field oscillations in sufficiently steep potentials~\cite{Turner:1983he,Poulin:2018dzj,Poulin:2018cxd,Agrawal:2019lmo}, or modified gravity~\cite{SanchezLopez:2023ixx}.

The impact of post-inflationary stiff phases on $\Omega_{\rm GW}^0(f)$ has been extensively studied~\cite{Dimopoulos:2022mce,Gouttenoire:2021wzu,Ahmad:2017itq,Co:2021lkc,Chen:2024roo,Brissenden:2025ass}, particularly in scenarios where the stiff equation of state arises from the dynamics of a canonical scalar field~\cite{Giovannini:1998bp,Giovannini:1999bh,Tashiro:2003qp}. However, the possibility of modifying the EoS through non-canonical kinetic terms has received comparatively less attention (see \textit{e.g.} Ref.~\cite{SanchezLopez:2023ixx,Chowdhury:2023opo}). In this work we explore this possibility by considering the Dirac--Born--Infeld (DBI) action~\cite{Silverstein:2003hf,Alishahiha:2004eh,Chen:2004gc,Chen:2005ad,Shandera:2006ax}, arising in string theory as the effective action describing the dynamics of a D3-brane moving through a warped throat region of a compact internal space. In this framework, the inflaton is identified with the position of the D3-brane and its kinetic term is non-canonical. We show that the DBI kinetic structure can reduce the EoS parameter relative to canonical kination, thereby softening the blue tilt of the GW spectra during stiff phases. We further investigate the resulting GW spectra numerically and explore the parameter space compatible with cosmological constraints, assessing the prospects for detection by future GW experiments.

The paper is organized as follows. In Sec.~\ref{sec:setup}, we present the general framework of $k$-inflation, at both the level of the background and the perturbations. We then specialize to the DBI action and analyze the post-inflationary dynamics and resulting energy-density spectrum of GWs. In Sec.~\ref{sec:numerics}, we describe the numerical approach to solving the system during and after inflation, and explore the parameter space of the model subject to cosmological constraints. We give our concluding remarks in Sec.~\ref{sec:conclusions}. In Appendix~\ref{Appendix}, we carry out an analytical investigation of constant-roll DBI backgrounds with evolving sound speed.

Throughout this work, we use natural units $c = \hbar = 1$ together with the reduced Planck mass $\m =(8\pi G)^{-1/2} = 2.44 \times 10^{18}$GeV, and adopt the ``mostly plus'' convention for the metric signature $(-,+,+,+)$.

\section{Setup} \label{sec:setup}
In this section we present the general theoretical framework under which our model is constructed, together with the central result: for a post-inflationary phase dominated by a kinetically-dominated scalar field governed by a DBI action, the EoS parameter is smaller than in canonical kination, leading to a softer blue tilt in the present-day density spectrum of GWs.

\subsection{Inflation}
\label{inflation}
Here, we provide an overview of $k$-inflation and refer the reader to Refs.~\cite{Armendariz-Picon:1999hyi,Garriga:1999vw,Seery:2005wm} for additional details. The starting point is the action
\begin{equation}
    S = \int \text{d}^4x\sqrt{-g}\left[\frac{\m^2}{2}R+P(X,\phi)\right],
\end{equation}
where $\phi$ is a scalar field playing the role of the inflaton, and $P=P(X,\phi)$ is a function of $\phi$ and
\begin{equation}
    X\equiv-\frac{1}{2}g^{\mu\nu}\partial_\mu\phi\partial_\nu\phi.
\end{equation}
The background dynamics are governed by the Friedmann and Raychaudhuri equations
\begin{equation}
    \label{eq:Friedmann}
    H^2=\frac{\rho}{3\m^2}=\frac{2X P_X  - P}{3\m^2},
\end{equation}
\begin{equation}
    \label{eq:2ndFriedmannP}
    \dot{H}=-\frac{\rho+p}{2\m^2}=-\frac{X P_X}{\m^2},
\end{equation}
together with the inflaton Euler--Lagrange equation
\begin{equation}
    \label{eq:EL}
    \frac{\text{d}}{\text{d}t}(a^3 P_X \dot{\phi})=a^3 P_\phi,
\end{equation}
where $X=\dot{\phi}^2/2$, $P_X \equiv \partial P/\partial X$ (and analogously for $P_{\phi}$ and higher order derivatives), $a$ is the scale factor, and $H=\dot a /a$ is the Hubble parameter. Eq.~\eqref{eq:EL} can also be obtained from the conservation of the energy-momentum tensor $\nabla_\mu T^{\mu\nu} = 0$, leading to the Klein--Gordon (KG) equation
\begin{equation}
    \label{eq:KGP}
    \ddot{\phi}\left(P_X+2XP_{XX}\right)+3H P_X \dot{\phi} + 2X P_{X\phi} - P_{\phi} = 0.
\end{equation}

Due to the modified kinetic structure of the action, the inflaton fluctuations travel with a sound speed given by
\begin{equation}
    c_\mathrm{s}^2=\frac{P_X}{2 X P_{XX} + P_X},
\end{equation}
which is different from unity unless $P_{XX}=0$. In general this is not equal to the EoS parameter
\begin{equation}
    w^{(\phi)} =\frac{p^{(\phi)}}{\rho^{(\phi)}}=\frac{P}{2X P_X - P}.
\end{equation}

It is convenient to define the Hubble flow parameters $\epsilon_n$, $\delta_n$ as
\begin{align}
    \label{eq:flowparams}
    \epsilon_{n+1}=\frac{\mathrm{d}\log \abs{\epsilon_n}}{\mathrm{d}N}\, ,\quad \delta_{n+1}=\frac{\mathrm{d}\log \abs{\delta _n}}{\mathrm{d}N}\,, 
\end{align}
with $\epsilon_0=H_\mathrm{ini}/H$ and $\delta_0=c_{_\mathrm{S},\,\mathrm{ini}}/c_{_\mathrm{S}}$. $N=\log{a/a_{\rm e}}$\footnote{Throughout this work, the super/sub-script ``e'' is used for quantities evaluated at the end of inflation.} is the number of e-folds relative to the end of inflation. With this convention, $N<0$ during inflation and $N>0$ during the post-inflationary evolution. When $\abs{\epsilon_{n\geq1}}\ll 1$ and $\abs{\delta_{n\geq1}}\ll 1$, the usual slow-roll approximations for the background and perturbations are admitted. 

Choosing the comoving gauge, in which $\delta\phi=0$, the single propagating scalar degree of freedom\footnote{The lapse and shift perturbations $\varphi$ and $B$ are not dynamical and can be eliminated using the $00$ and $0i$ Einstein equations.} can be described by the comoving curvature perturbation $\mathcal{R}$. Tensor perturbations are described by the gauge invariant, transverse and traceless metric perturbation $h_{ij}$, satisfying $h^i{}_i=\partial_ih^{ij}=0$. The line element is then written as\footnote{We do not consider vector perturbations since they quickly decay during inflation.}
\begin{align}
    \text{d}s^2 &= a^2\Big[-(1+\varphi)\text{d}\eta^2 + 2\partial_i B \text{d}\eta\text{d}x^i +\nonumber\\
    &+(1+2\mathcal{R})\delta_{ij}\text{d}x^i\text{d}x^j + h_{ij}\text{d}x^i\text{d}x^j\Big],
\end{align}
eventually leading to the second order action~\cite{Chen:2006nt}
\begin{align}
    \label{eq:secondorderactionR}
    S&=\frac{1}{2}\int \text{d}\eta\text{d}^3x\, z^2 \left[(\R')^2-c_{\rm s}^2(\partial \R)^2\right]=\nonumber\\
    &=\frac{1}{2}\int \text{d}\eta\text{d}^3x\,\left[(v')^2+\frac{z''}{z}v^2-c_{\rm s}^2(\partial v)^2\right],
\end{align}
where primes denote derivatives with respect to conformal time. Here we have defined 
\begin{equation}
    z = \frac{\m a \sqrt{2\epsilon_1}}{c_{\rm s}}
\end{equation}
together with the Mukhanov--Sasaki (MS) variable
\begin{equation}
    \label{eq:MSvariable}
    v = z \mathcal{R}.
\end{equation}
We promote the MS variable to a quantum operator $v\to\hat{v}$ as
\begin{equation}
    \hat{v}(\eta, \bm{x}) = \int \frac{\text{d}^3\bm{k}}{(2\pi)^3}\left[v_k(\eta)\hat{a}_{\bm{k}}e^{i \bm{k}\cdot\bm{x}} + v_k^{*}(\eta)\hat{a}_{\bm{k}}^{\dagger}e^{-i \bm{k}\cdot\bm{x}}\right],
\end{equation}
where $\hat{a}_{\bm{k}}$ and $\hat{a}_{\bm{k}}^{\dagger}$ are annihilation and creation operators, respectively, that satisfy canonical commutation relations $[\hat{a}_{\bm{k}},\hat{a}_{\bm{p}}^{\dagger}]=(2\pi)^3\delta^{(3)}(\bm{k} - \bm{p})$. The mode functions $v_k(\eta)$ follow the MS equation, derived from Eq.~\eqref{eq:secondorderactionR},
\begin{equation}
    \label{eq:veom}
    v_k''+\omega_k^2(\eta)v_k=0,
\end{equation}
where
\begin{align}
    &\omega_k^2(\eta) = c_\mathrm{s}^2k^2-\frac{z''}{z}=\nonumber\\
    &= c_\mathrm{s}^2k^2-\mathcal{H}^2\Bigg[2-\epsilon_1+\frac32 \epsilon_2+\frac14 \epsilon_2^2 -\frac12 \epsilon_1 \epsilon_2 +\frac12 \epsilon_2 \epsilon_3+\nonumber\\
    &+\left(3-\epsilon_1+\epsilon_2\right)\delta_1+\delta_1^2
    +\delta_1 \delta_2\Bigg].
\end{align}

For modes sufficiently deep inside the sound horizon $c_{\rm s}^2k^2\gg\abs{z''/z}$, their equation of motion becomes
\begin{equation}
    v_k'' + c_{\rm s}^2k^2v_k=0.
\end{equation}
As long as $\omega_k(\eta)$ varies adiabatically
\begin{align}
    &\abs{\frac{3(c_{\rm s}')^2}{4c_{\rm s}^4 k^2}-\frac{c_{\rm s}''}{2c_{\rm s}^3k^2}} =\nonumber\\
    &=\frac{1}{4}\left(\frac{\mathcal{H}}{c_{\rm s}k}\right)^2\abs{\delta_1^2 + 2\delta_1(1+\delta_2-\epsilon_1)}\ll1,
\end{align}
the WKB method may be employed and the adiabatic vacuum imposed~\cite{Martin:2002vn}
\begin{equation}
    \label{eq:BDic}
    \lim_{\abs{c_{\rm s}k/\mathcal{H}}\to \infty}v_{k}(\eta) = \frac{1}{\sqrt{2 c_{\rm s}(\eta)k}}e^{-i k\int^\eta \text{d}\Tilde{\eta} c_{\rm s}(\tilde{\eta})}.
\end{equation}
In practice, for smoothly varying Hubble flow parameters and $\abs{\delta_2}$ order unity at most, the condition $\abs{\delta_1}\lesssim 1$ is sufficient to ensure sub-horizon adiabatic evolution, since for such modes $c_{\rm s}k\gg \mathcal{H}$.

Finally, the variance of the quantum comoving curvature perturbation at the end of inflation can be related to the solutions of Eq.~\eqref{eq:veom} as
\begin{equation}
    \langle\hat{\mathcal{R}}(\eta_{\rm e},\bm{x})^2\rangle = \int \text{d}\log k\, \mathcal{P}_{\mathcal{R}}(k),
\end{equation}
where $\eta_{\rm e}$ is the conformal time at the end of inflation, defined via $\epsilon_1(\eta_{\rm e})=1$. Here, the dimensionless power spectrum is given by
\begin{equation}
    \label{eq:Rpowerspectrum}
    \mathcal{P}_{\mathcal{R}}(k) = \frac{k^3}{2\pi^2}\Bigg|\frac{v_k(\eta_{\rm e})}{z(\eta_{\rm e})}\Bigg|^2.
\end{equation}
It is useful to parametrize the power spectrum through a Taylor expansion of its logarithm around the CMB pivot scale $k_{*}=0.05$Mpc$^{-1}$ 
\begin{equation}
    \mathcal{P}_{\mathcal{R}}(k)=A_{\rm s}\left(\frac{k}{k_{*}}\right)^{n_{\rm s}-1 + \frac{\alpha_{\rm s}}{2}\log\left(k/k_{*}\right)},
\end{equation}
such that
\begin{equation}
\begin{gathered}
    A_{\rm s} = \mathcal{P}_{\mathcal{R}}(k_{*}),
    \quad
    n_{\rm s}-1=\frac{\text{d}\log \mathcal{P}_{\mathcal{R}}}{\text{d}\log k}\Bigg|_{k_{*}}, \\
    \alpha_{\rm s} = \frac{\text{d}n_{\rm s}}{\text{d}\log k}\Bigg|_{k_{*}},
\end{gathered}
\end{equation}
where $A_{\rm s}$ is the amplitude of the power spectrum, $n_{\rm s}$ the scalar spectral index, and $\alpha_{\rm s}$ its running.

The treatment of tensor perturbations proceeds analogously to the scalar sector. Since in our theory the matter sector is minimally coupled to gravity, tensor perturbations retain a canonical kinetic structure and propagate at the speed of light. Consequently, their evolution depends on the non-canonical dynamics only through the background.\footnote{This contrasts with extensions involving DBI scalar fields coupled to massive gravity, where the tensor dispersion relation can be modified~\cite{Kazempour:2022giy}.} The second order action for the tensor perturbations reads
\begin{align}
    \label{eq:secondorderactiontensors}
    S &= \frac{\m^2}{4}\sum_{s=+,\cross}\int \text{d}\eta\text{d}^3x\,a^2\left[(h^s{}')^2 - (\partial h^s)^2\right]=\nonumber\\
    &=\frac{1}{2}\sum_{s=+,\cross}\int \text{d}\eta\text{d}^3x\,\left[(u^s{}')^2 + \frac{a''}{a}(u^s)^2 - (\partial u^s)^2\right],
\end{align}
where we have introduced the canonically normalized tensor variable
\begin{equation}
    u^s = \frac{\m a}{\sqrt{2}}h^s.
\end{equation}
Here, $s$ indexes the two different gravitational wave polarizations. The Fourier components of $h^s$ are related to the tensor modes $h_{ij}$ via
\begin{equation}
    h_{ij}(\eta,\bm{x}) = \sum_{s=+,\cross}\int\frac{\text{d}^3k}{(2\pi)^3}p_{ij}^s(\bm{k})h_{\bm k}^s(\eta) e^{i\bm{k}\cdot\bm{x}},
\end{equation}
where $p_{ij}^s(\bm{k})$ is the polarization tensor satisfying $p_{ii}^s=k^ip_{ij}^s=0$, together with the normalization $p_{ij}^s (\bm{k})p_{ij}^{s'}(\bm{k})=2\delta_{ss'}$.

We promote the canonically normalized variable to a quantum operator $u^s\to\hat{u}^s$ as
\begin{equation}
    \hat{u}^s(\eta, \bm{x}) = \int \frac{\text{d}^3\bm{k}}{(2\pi)^3}\left[u^s_k(\eta)\hat{a}^s_{\bm{k}}e^{i \bm{k}\cdot\bm{x}} + u_k^s{}^{*}(\eta)(\hat{a}_{\bm{k}}^s)^{\dagger}e^{-i \bm{k}\cdot\bm{x}}\right],
\end{equation}
where $\hat{a}^s_{\bm{k}}$ and $(\hat{a}_{\bm{k}}^s)^{\dagger}$ are annihilation and creation operators, respectively, that satisfy canonical commutation relations $[\hat{a}^s_{\bm{k}},(\hat{a}_{\bm{p}}^{s'})^{\dagger}]=(2\pi)^3\delta^{ss'}\delta^{(3)}(\bm{k} - \bm{p})$. The mode functions $u^s_k(\eta)$ obey the following equation of motion
\begin{equation}
    \label{eq:veomT}
    (u^s_k)''+\left(k^2 - \frac{a''}{a}\right)u^s_k=0.
\end{equation}

Tensor modes are initialized in the Bunch--Davies vacuum
\begin{equation}
    \label{eq:BDicT}
    \lim_{\abs{k/\mathcal{H}}\to \infty}u^s_k(\eta) = \frac{1}{\sqrt{2 k}}e^{-i k \eta},
\end{equation}
while on super-horizon scales $k^2\ll a''/a$ the solution reads
\begin{equation}
    h^s_k(\eta) = C_1 + C_2 \int^\eta\frac{\text{d}\Tilde{\eta}}{a^2(\Tilde{\eta})}.
\end{equation}
During inflation, the decaying mode rapidly becomes negligible, showing that tensor modes are conserved on super-horizon scales. This behaviour depends solely on $\epsilon_1$, rather than on $\epsilon_1$, $\epsilon_2$, and $\delta_1$, as for the scalar modes. 

The variance of the quantum tensor modes at the end of inflation can be related to the solutions of Eq.~\eqref{eq:veomT} as
\begin{equation}
    \langle \hat{h}_{ij}(\eta_{\rm e},\bm{x})\hat{h}^{ij}(\eta_{\rm e},\bm{x})\rangle = \int \text{d}\log k\,\mathcal{P}_h(k),
\end{equation}
where
\begin{equation}
    \mathcal{P}_h(k)=\sum_{s=+,\cross}\frac{k^3}{\pi^2}\abs{h_k^s(\eta_{\rm e})}^2 = \frac{4 k^3}{\m^2 \pi^2 a^2(\eta_{\rm e})}\abs{u^s_k(\eta_{\rm e})}^2,
\end{equation}
where we have used that both polarization modes satisfy the same equation of motion in the last step. As for the scalar case, the tensor power spectrum is parametrized through a Taylor expansion of its logarithm around the CMB pivot scale $k_{*}$ as
\begin{equation}
    \mathcal{P}_h(k) = rA_{\rm s}\left(\frac{k}{k_{*}}\right)^{n_{\rm t}},
\end{equation}
such that
\begin{align}
    r = \frac{\mathcal{P}_h(k_{*})}{\mathcal{P}_{\mathcal{R}}(k_{*})},\quad n_{\rm t} =\frac{\text{d}\log \mathcal{P}_{h}}{\text{d}\log k}\Bigg|_{k_{*}},
\end{align}
where $r$ is the tensor-to-scalar ratio and $n_{\rm t}$ is the tensor spectral index.

Instead of the power spectrum, late-time observations are often discussed in terms of the energy density of the tensor modes,
\begin{equation}
    \label{eq:rho_h}
    \langle\hat\rho_h(\eta)\rangle = \int \text{d}\log k\, \rho_k^h(\eta),
\end{equation}
where
\begin{equation}
    \label{eq:rho_h_k}
    \rho_k^h(\eta) = \frac{\m^2}{2} \sum_{s=+,\cross} \frac{k^3}{4\pi^2 a^2(\eta)} \left[|h_k^s{}'(\eta)|^2 + k^2|h_k^s(\eta)|^2  \right].
\end{equation}
At super-horizon scales, when the modes are frozen, the derivative term becomes subleading, and Eq.~\eqref{eq:rho_h_k} simplifies to
\begin{equation}
    \label{eq:rho_h_k_sh}
    \rho_k^h(\eta) = \frac{\m^2}{8}\frac{k^2}{a^2(\eta)}\mathcal{P}_h(k).
\end{equation}

Super-horizon modes stay frozen also after the end of inflation, thawing only upon horizon re-entry. Inside the horizon, the $u_k^s$ variables start to oscillate with a constant amplitude, so the corresponding $h_k^s$ amplitude decays as $1/a$. Both $h^s_k{}'$ and $h_k^s$ terms contribute equally to the energy density which scales like radiation, $\rho^h_k \propto \frac{k^2}{a^2}|h^s_k|^2 \propto a^{-4}$.

\subsection{The DBI model}
In this work, we consider the DBI~\cite{Silverstein:2003hf,Alishahiha:2004eh} action
\begin{equation}
    P(X,\phi)=\frac{1}{f(\phi)}\left(1-\sqrt{1- 2 f(\phi)X}\right) - V(\phi),
    \label{eq:P_DBI}
\end{equation}
where $f(\phi)>0$ is the warp factor, corresponding to the inverse D3-brane tension, and $V(\phi)$ is the inflaton potential. Since $2 f(\phi) X<1 $ for the square-root to be real, the DBI action admits a series expansion in powers of the kinetic term. At leading order, in the non-relativistic limit $2f(\phi)X\ll1$, one recovers the canonical scalar field Lagrangian $P(X,\phi)=X-V(\phi)$. When $2f(\phi)X$ becomes sufficiently large, higher order terms in the expansion can become significant, leading to deviations from the canonical dynamics.

It will be useful to define the Lorentz factor, related to the speed of sound via
\begin{equation}
    \label{eq:gamma}
    \gamma =\frac{1}{\sqrt{1 - 2 f(\phi)X}} = \frac{1}{c_{\rm s}}.
\end{equation}
It satisfies $\gamma \geq 1$. The energy density and pressure are readily obtained to be
\begin{equation}
    \label{eq:rhoPhiDBI}
    \rho^{(\phi)} = \frac{\gamma -1 }{f} + V = \frac{\gamma^2}{1+\gamma}\dot{\phi}^2+V,
\end{equation}
and
\begin{equation}
    p^{(\phi)} = \frac{\gamma -1}{\gamma f}-V= \frac{\gamma}{1+\gamma}\dot{\phi}^2-V.
\end{equation}

For future convenience, we interpret the homogeneous scalar field $\phi=\phi(t)$ as the time variable, such that the Hubble parameter becomes $H(\phi)$, in the so-called Hamilton--Jacobi formalism~\cite{Kinney:1997ne}. The Friedmann and Raychaudhuri equations read
\begin{equation}
    \label{eq:1stHamJac}
    3\m^2H^2=\frac{\gamma^2}{1+\gamma}\dot{\phi}^2+V,
\end{equation}
\begin{equation}
    \label{eq:2ndHamJac}
    H_{\phi} = - \frac{\gamma \dot{\phi}}{2\m^2},
\end{equation}
together with
\begin{equation}
    \gamma = \sqrt{1 + 4\m^4 f H_{\phi}^2}.
\end{equation}
These equations are valid when the inflaton dominates the energy density, both during inflation and during the post-inflationary stiff phase, as long as the radiation component remains negligible. The KG equation for the homogeneous DBI field reads
\begin{equation}
    \label{eq:KG-DBI}
    \ddot{\phi} + \frac{3 H \dot{\phi}}{\gamma^2} +\frac{f_{\phi}}{2f^2 \gamma^3}\left(\gamma-1\right)^2(\gamma+2) + \frac{V_\phi}{\gamma^3}=0.
\end{equation}

\begin{figure}
    \centering
    \includegraphics[width=0.8\columnwidth]{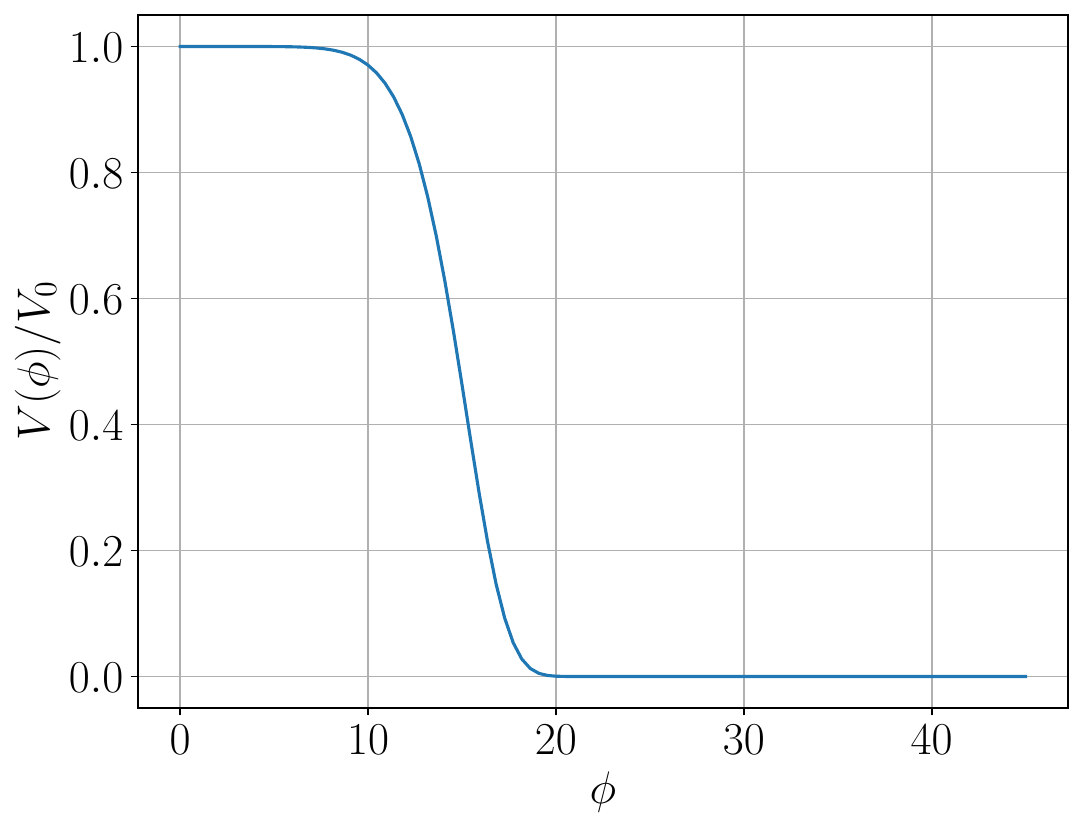}
    \caption{Inflaton potential \eqref{eq:potential} for reference parameter values~\eqref{eq:parametervalues} $n=8$ and $\lambda=3\times10^{-10}$.}
    \label{fig:potential}
\end{figure}

We fix the potential to be (see Fig.~\ref{fig:potential})
\begin{equation}
    \label{eq:potential}
    V(\phi) = V_0 e^{-\lambda \phi^n / \m^n},
\end{equation}
which was proposed in Ref.~\cite{Geng:2017mic} as a quintessential inflation potential, where the inflaton field survives until the present day to become dark energy. We emphasize that in the present work we simply use Eq.~\eqref{eq:potential} as a concrete realization of an inflationary period followed by a kinetically dominated epoch, without imposing any dark energy constraints. In our setup the field starts rolling near the small-$\phi$ plateau of Fig.~\ref{fig:potential} with a positive value $\phi>0$, along the negative slope branch of the potential $V_\phi<0$, and continues towards increasing values of $\phi$ such that $\dot{\phi}>0$.

\subsection{Post-inflationary DBI dynamics} \label{sec:postinflation}
After inflation ends, the inflaton keeps rolling along the runaway potential, while a radiation component is produced by an alternative reheating mechanism, such as gravitational reheating~\cite{Ford:1986sy,Lankinen:2019ifa,Chun:2009yu,deHaro:2023gho} or instant preheating~\cite{Felder:1998vq,Felder:1999pv,Ellis:2020krl}.\footnote{Additional mechanisms include Ricci reheating~\cite{Opferkuch:2019zbd,Laverda:2023uqv,Dimopoulos:2018wfg}, curvaton reheating~\cite{Feng:2002nb}, and black hole evaporation~\cite{Dalianis:2021dbs}.} To keep the analysis applicable to a broad range of reheating mechanisms, we parametrize the energy density of radiation via its value at the end of inflation, $\rho_\ra^{\rm e}$, without specifying the details of its production. Assuming the radiation fluid is separately conserved, its energy density reads
\begin{equation}
    \label{eq:rhor}
    \rho_\ra(N) = \rho_\ra^{\rm e} e^{-4N}.
\end{equation}

Choosing the normalization $a_0=1$,\footnote{Here and below, the super/sub-script ``0'' is used for quantities evaluated at present.} the scale factor at the end of inflation is fixed given $\rho_\ra^{\rm e}$
\begin{equation}
    \label{eq:ae}
    a_{\rm e} = \left(\frac{\rho_\ra^0}{\rho_\ra^{\rm e}}\right)^{1/4}.
\end{equation}
The energy density of radiation at present is given by
\begin{equation}
    \rho_\ra^0 = 9.63\times 10^{-125}\m^4,
\end{equation}
where we have used $N_{\rm eff}=3.044$~\cite{Akita:2020szl,Froustey:2020mcq,Bennett:2020zkv,Drewes:2024wbw} and $T_{\rm CMB} = 2.7255\rm{K}$~\cite{Baumann:2022mni}. For simplicity, we neglect changes in the effective number of relativistic degrees of freedom $g_{*}(T)$ and $g_{*s}(T)$. For tensor modes re-entering the horizon at sufficiently high temperatures, the correction reduces approximately to an overall normalization. Since $\rho_\ra^{\rm e}$ is treated as a phenomenological free parameter, it can be chosen to partly absorb such a correction. For temperatures approximately below $10^3$GeV, changes in $g_{*}(T)$ and $g_{*s}(T)$ introduce some scale dependence in the GW spectrum~\cite{Watanabe:2006qe}, but this would not change the qualitative conclusions of our analysis, mainly focused on the post-inflationary stiff phase.

Including the radiation component, the Friedmann and Raychaudhuri equations are modified as
\begin{equation}
    \label{eq:1stFriedmannpostinfl}
    H^2=\frac{1}{3\m^2}\left(\frac{\gamma-1}{f}+V+\rho_\ra\right),
\end{equation}
\begin{equation}
    \label{eq:2ndFriedmannpostinfl}
    \dot{H} = -\frac{\gamma\dot{\phi}^2+\frac{4}{3}\rho_\ra}{2\m^2}.
\end{equation}
The KG equation remains given by Eq.~\eqref{eq:KG-DBI}. The total EoS parameter is
\begin{equation}
    \label{eq:wtot}
    w_{\rm tot} = \frac{p^{(\phi)}+\frac{1}{3}\rho_\ra}{\rho^{(\phi)}+\rho_\ra}.
\end{equation}

Regarding $\phi$, after inflation ends, it exits slow-roll and starts rolling down a progressively steeper potential, becoming kinetically dominated
\begin{align}
    X\gg V.
\end{align}
Since $\gamma\geq 1$, this condition ensures that $\gamma^2\dot{\phi}^2/(1+\gamma)\gg V$ and 
$\gamma\dot{\phi}^2/(1+\gamma)\gg V$, so that the EoS parameter simplifies as
\begin{equation}
    \label{eq:wkineticdomination}
    w^{(\phi)} = \frac{\frac{\gamma}{1+\gamma}\dot{\phi}^2-V}{\frac{\gamma^2}{1+\gamma}\dot{\phi}^2+V}\overset{X\gg V}{\simeq} \frac{1}{\gamma} = \sqrt{1 - 2f(\phi) X}.
\end{equation}
We notice in this regime $w^{(\phi)} \simeq c_s$. Using basic relations, the DBI Lagrangian can be expressed as 
\begin{equation}
    P(X,\phi) = \frac{\gamma}{\gamma + 1}2X -V \overset{X\gg V}{\simeq} \frac{\gamma}{\gamma + 1}2X.
\end{equation}
This Lagrangian does not belong to the class $P\propto X^n$, for which $w^{(\phi)}=c_{\rm s}^2 = 1/(2n - 1)$, since $\gamma$ depends on both $\phi$ and $X$. The fact that $w^{(\phi)} \simeq c_s$ in the kinetically dominated regime is due to the specific DBI structure.

Since $\infty>\gamma>1$ for $f(\phi)>0$ and $X>0$, Eq.~\eqref{eq:wkineticdomination} implies that $0<w^{(\phi)}<1$ in the kinetically dominated regime. In the canonical limit $f(\phi)X\to 0$, one recovers the kination result $w^{(\phi)}=1$. Therefore, in general, \textit{the kinetic structure provided by the DBI action makes the EoS parameter smaller than one}. Obtaining a stiff phase $1/3<w^{(\phi)}< 1$ requires $0 < f(\phi) X < \frac{4}{9}$. In this regime, the density of the inflaton decays faster than that of radiation. We define as ``reheating'' the time $N_{\rm reh}$ at which $\rho_\phi(N_{\rm reh})=\rho_\ra(N_{\rm reh})$.

Although strictly $w^{(\phi)}<1$, its time evolution and magnitude of the departure from the canonical limit are model dependent. In order to identify a warp factor capable of producing a substantial effect on the post-inflationary dynamics, let us find what $f(\phi)$ is required to have constant $w^{(\phi)}$ (or constant $f(\phi)X$). The following estimates rely on the assumptions of scalar/kinetic domination; we therefore expect departures from the full numerical evolution, which properly accounts for the transition out of inflation and for the increasing importance of the radiation component as reheating is approached (see the top panel of Fig.~\ref{fig:w}). In Appendix~\ref{Appendix}, we extend the discussion to stiff backgrounds with evolving sound speed.

Scalar/kinetic domination allows us to neglect the radiation component and use Eq.~\eqref{eq:gamma} to express Eq.~\eqref{eq:2ndHamJac} as
\begin{equation}
    \label{eq:fHamJac}
    f = \frac{\gamma^2 - 1}{4\m^4H_\phi^2}.
\end{equation}
Then, taking the kinetic domination limit in Eq.~\eqref{eq:1stHamJac} gives 
\begin{equation}
    \label{eq:fHamJabKinetic}
    f\overset{X\gg V}{\simeq} \frac{\gamma-1}{3\m^2 H^2}.
\end{equation}
Equating Eqs.~\eqref{eq:fHamJac} and~\eqref{eq:fHamJabKinetic} and integrating we find
\begin{equation}
    H(\phi) = H_i \exp\left\{-\int^\phi \text{d}\Tilde{\phi}\frac{\sqrt{3\left(\gamma(\Tilde{\phi})+1\right)}}{2\m}\right\},
\end{equation}
where we choose the negative sign of the square root such that $\dot{\phi}>0$, consistently with the negative slope of the potential. This equation is valid for any scalar-kinetically-dominated phase. Imposing the extra assumption of a constant EoS parameter gives
\begin{equation}
    \label{eq:constantH}
    H = H_i \exp\left\{-\frac{\phi\sqrt{3\left(\gamma+1\right)}}{2\m}\right\}.
\end{equation}
Finally, plugging this equation back in Eq.~\eqref{eq:fHamJabKinetic}, the required warp factor is given by
\begin{equation}
    \label{eq:f}
    f(\phi) = f_0 e^{\kappa \phi/\m},
\end{equation}
where
\begin{equation}
    \label{eq:kappaintermsofw}
    \kappa = \sqrt{3\left(\frac{1}{w^{(\phi)}}+1\right)}.
\end{equation}
We will use the exponential warp factor \eqref{eq:f} in our numerical analysis below. It was previously considered in the context of DBI quintessence~\cite{Martin:2008xw}, dynamical-systems analyses of DBI cosmology~\cite{Copeland:2010jt} and DBI power-law inflation~\cite{Amani:2018ueu,Tsujikawa:2013ila}. In order to satisfy $1/3<w^{(\phi)}< 1$, we restrict our analysis to
\begin{equation} \label{eq:kappa_limits}
    \sqrt{6} < \kappa < \sqrt{12} .
\end{equation}

\subsection{Gravitational wave spectrum}\label{GWs}
Above, we computed the primordial gravitational wave spectrum produced during inflation. The post-inflationary dynamics leave a mark on the spectrum, generating a signature that can be probed observationally today.

As noted at the end of section~\ref{inflation}, tensor perturbations stay frozen until horizon re-entry, after which their amplitude scales as $a^{-1}$, and their energy density scales as radiation. Starting from Eq.~\eqref{eq:rho_h_k_sh} and taking into account the correct sub-horizon redshifting, we thus obtain the $k$ mode GW energy fraction today as
\begin{equation}
    \label{eq:rho_h_k_today}
    \Omega_\mathrm{GW}^0(k) = \frac{\rho_{k}^h(\eta_0)}{\rho_0} \simeq \frac{\m^2 H_k^2}{8\rho_0}\mathcal{P}_h(k) \left(\frac{a_k}{a_0}\right)^4,
\end{equation}
where $H_k$ and $a_k$ are the Hubble parameter and scale factor at re-entry ($k=a_k H_k$), and $\rho_0 = 3\m^2 H_0^2 = 1.04 \times 10^{-120}\m^4$ is the energy density today, computed from the Hubble parameter value $H_0 = 67.36$ km/s/Mpc \cite{Planck:2018vyg}. $H_k$ and $a_k$ depend on the stiff-era dynamics in a non-trivial way.

We emphasize that Eq.~\eqref{eq:rho_h_k_today} has been obtained assuming that all tensor modes are frozen before horizon re-entry. While this is an excellent approximation for most $k$, modes that left the horizon shortly before the end of inflation did not have time to completely freeze. Therefore, some discrepancy is expected in the UV part of $\Omega_\mathrm{GW}^0(k)$.

In the next section, we will solve $a_k$ and $H_k$ numerically to obtain $\Omega_\mathrm{GW}^0(k)$, but first, let us study the slope of $\Omega_\mathrm{GW}^0(k)$ analytically. Using the Friedmann and Raychaudhuri equations, we obtain the intermediate result, valid irrespective of the energy content of the Universe,
\begin{equation}
    \label{eq:dlnH_dlna}
    \frac{\dd \log H}{\dd \log a} = -\frac{3}{2}(1+w_{\rm tot}),
\end{equation}
where $w_{\rm tot}=p/\rho$ is the total EoS, given by Eq.~\ref{eq:wtot} in our model. Using the horizon re-entry condition, we then find
\begin{equation} \label{eq:Omega_GW_k_derivative}
\begin{aligned}
    \frac{\dd \log \Omega_\text{GW}^0(k)}{\dd \log k}
    &= \frac{\dd \log \mathcal{P}_h(k)}{\dd \log k} + \frac{\dd \log (H_k^2 a_k^4)}{\dd \log (a_k H_k)} \\
    &= n_{\rm t}(k) + \frac{2(3w_k - 1)}{3w_k+1},
\end{aligned}
\end{equation}
where $w_k$ is again evaluated at re-entry. For approximately scale invariant primordial perturbations ($n_{\rm t}\simeq0$), a period of canonical kination $w_k=1$ leads to a GW spectrum with slope equal to unity. However, since the DBI kinetic structure makes $1/3 < w_k<1$, we can conclude from Eq.~\eqref{eq:Omega_GW_k_derivative} that \textit{it softens the characteristic blue tilt of canonical kination, providing a GW spectral slope between zero and one during the stiff phase.}

Outside of the stiff era, during the transition out of inflation with $-1/3<w_k<1/3$, the GW slope is negative, while during the radiation dominated era with $w_k=1/3$, it vanishes. In our setup, $w_{\rm tot}$ grows from inflationary negative values to stiff positive ones, leaving behind a local maximum in the GW spectrum at the point where it first crosses $1/3$. The constant positive slope corresponding to the stiff period eventually turns into a plateau at lower frequencies when radiation becomes dominant. See the bottom panel of Fig.~\ref{fig:w} for an example of this behaviour.

\section{Numerical Analysis} \label{sec:numerics}
In this section, we study the model numerically, solving the background and perturbation equations during inflation, together with the post-inflationary evolution, to obtain the resulting GW spectra. We then explore the parameter space compatible with cosmological constraints and assess the prospects for detection by future GW experiments.

\subsection{Inflation}
\begin{figure*}
    \centering
    \includegraphics[width=\picwidth]{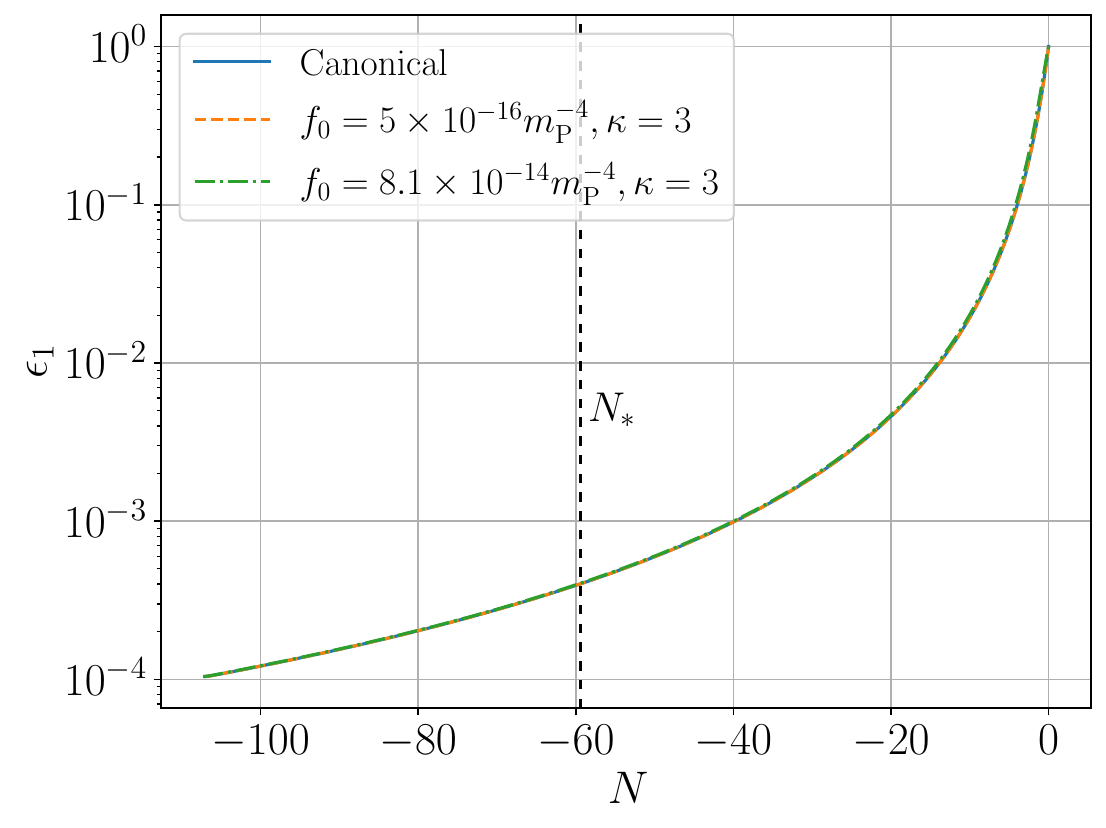}
    \picgap
    \includegraphics[width=\picwidth]{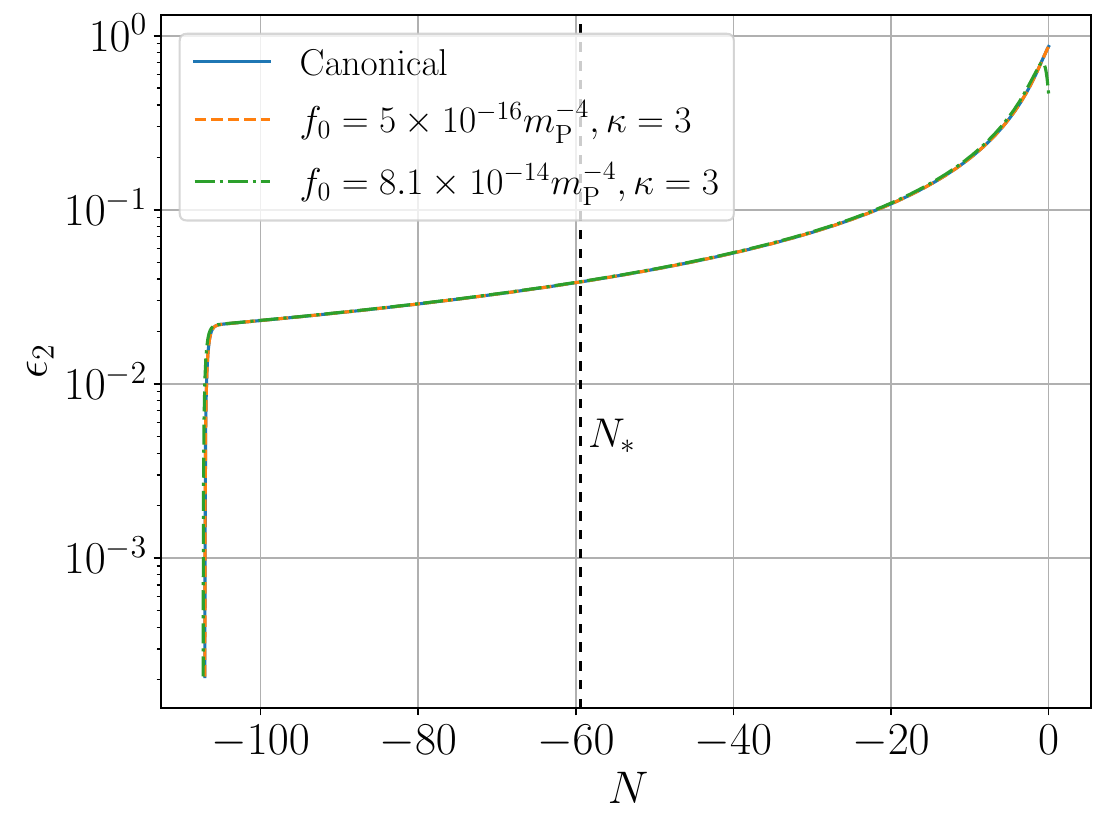}
    \includegraphics[width=\picwidth]{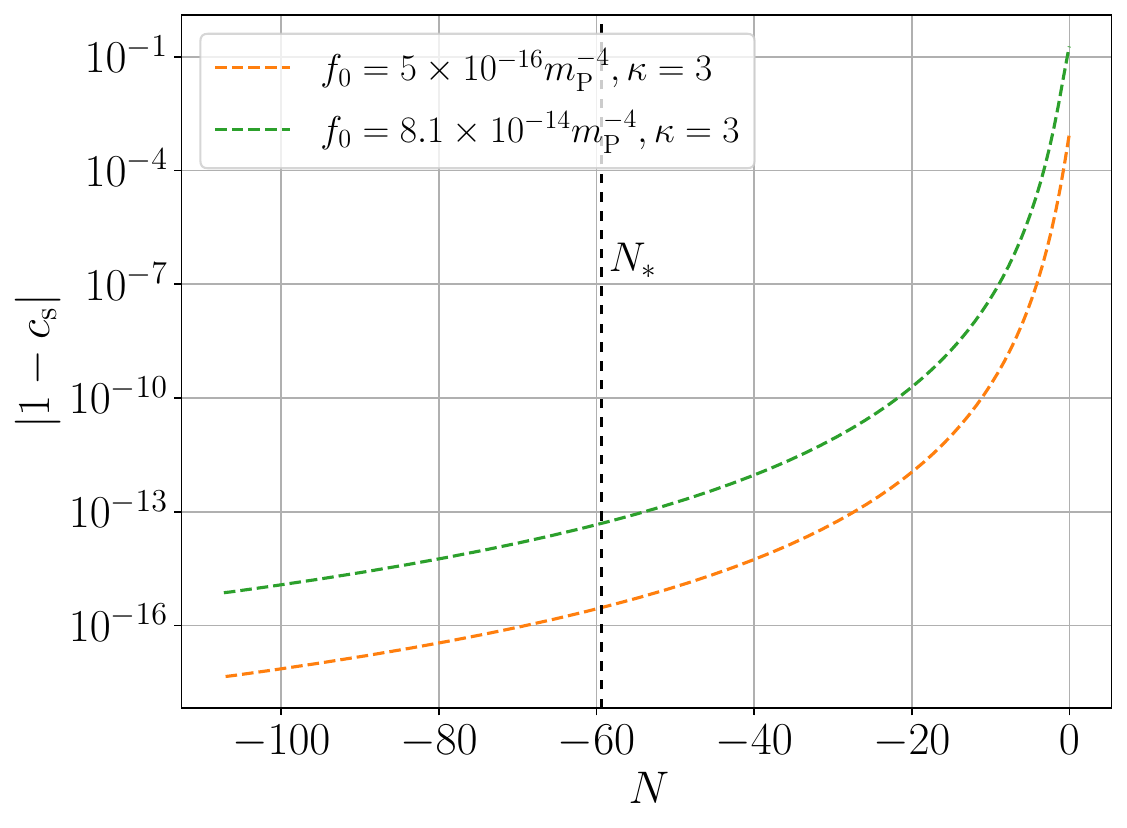}
    \picgap
    \includegraphics[width=\picwidth]{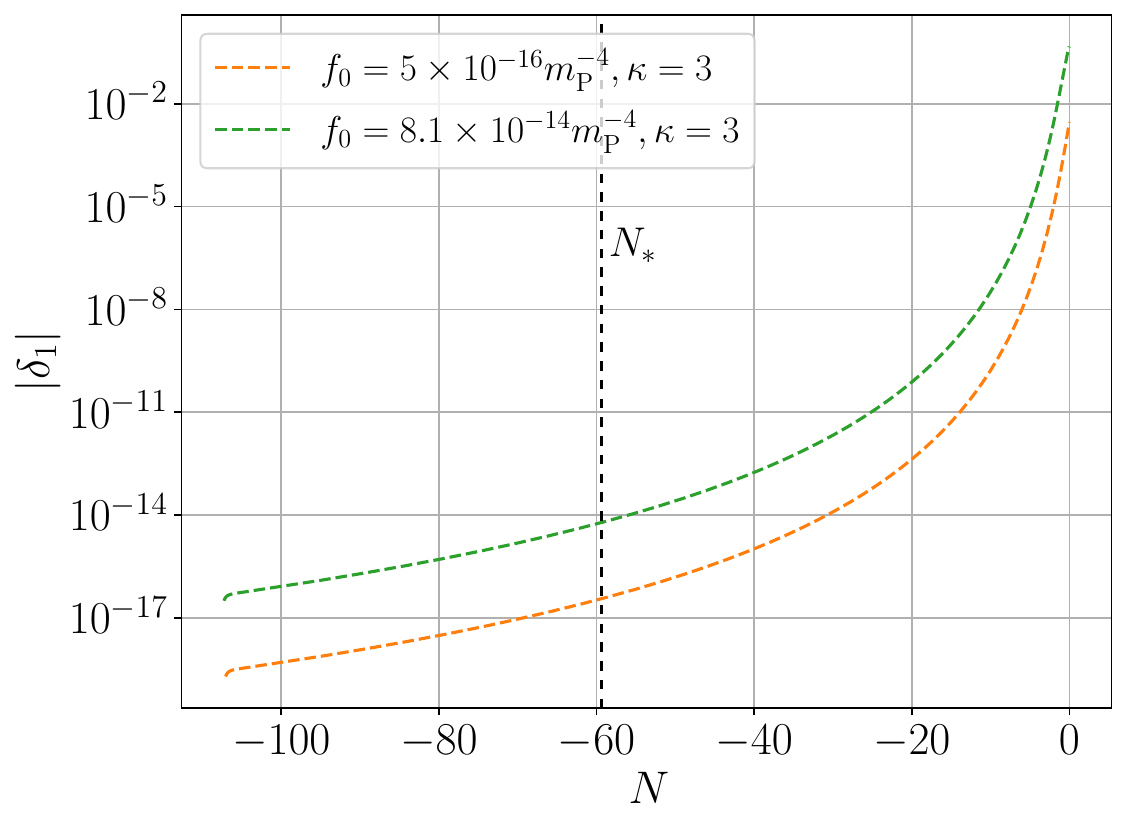}
    \caption{$\epsilon_1$ (top left), $\epsilon_2$ (top right), $\abs{1-c_{\rm s}}$ (bottom left), and $\abs{\delta_1}$ (bottom right) as a function of the number of e-folds before the end of inflation. In blue the dynamics is governed by a canonical action, while in dashed it is governed by a DBI action with an exponential warp factor~\eqref{eq:f}, with $\{f_0\m^4,\kappa\}=\{5\times 10^{-16}, 3\}$ (orange) and $\{f_0\m^4,\kappa\}=\{8.1\times 10^{-14}, 3\}$ (green). In the top panels, the lines overlap for most $N$. $N_{*}$ is the time at which the pivot scale left the horizon.}
    \label{fig:hubbleFlow}
\end{figure*}

The homogeneous background is numerically integrated by using the full Raychaudhuri~\eqref{eq:2ndFriedmannP} and KG~\eqref{eq:KG-DBI} equations, without imposing the slow-roll approximation during the evolution. Initial conditions are chosen such that the scalar field starts close to the slow-roll attractor, with an initial field value given by $\phi_i = 9.3\m$. In Fig.~\ref{fig:hubbleFlow}, the evolution of $\epsilon_1$, $\epsilon_2$, $\abs{c_{\rm s}-1}$ and $\abs{\delta_1}$ is shown for some representative cases, all compatible with the CMB constraints discussed below. Choosing a small enough value for $f_0$ (orange curves), the effect of the warp factor on the dynamics is negligible, with results nearly identical to the canonical case (blue curves) throughout inflation. Choosing a larger value for $f_0$ (green curves), the dynamics around the pivot scale remain essentially unaffected, while small deviations in $\epsilon_2$ become apparent near the end of inflation, together with increased departures of $c_{\rm s}$ from unity. For all cases, $\abs{\delta_1}\ll1$ during most of the evolution, supporting the adiabatic vacuum initialization for modes sufficiently deep inside the horizon.

The inflationary scalar and tensor power spectra are computed by solving Eqs.~\eqref{eq:veom} and~\eqref{eq:veomT} numerically on the resulting background. Integration for each mode starts sufficiently deep in the horizon, with $c_{\rm s}k\gtrsim100aH$ for scalar modes and $k\gtrsim100aH$ for tensor modes. The former are initialized in the adiabatic vacuum~\eqref{eq:BDic} and the latter are initialized in the Bunch--Davies vacuum~\eqref{eq:BDicT}. Evolution is integrated until the end of inflation.

\begin{figure}[ht] 
     \centering
     \begin{subfigure}[b]{0.4\textwidth}
         \centering
         \includegraphics[width=\textwidth]{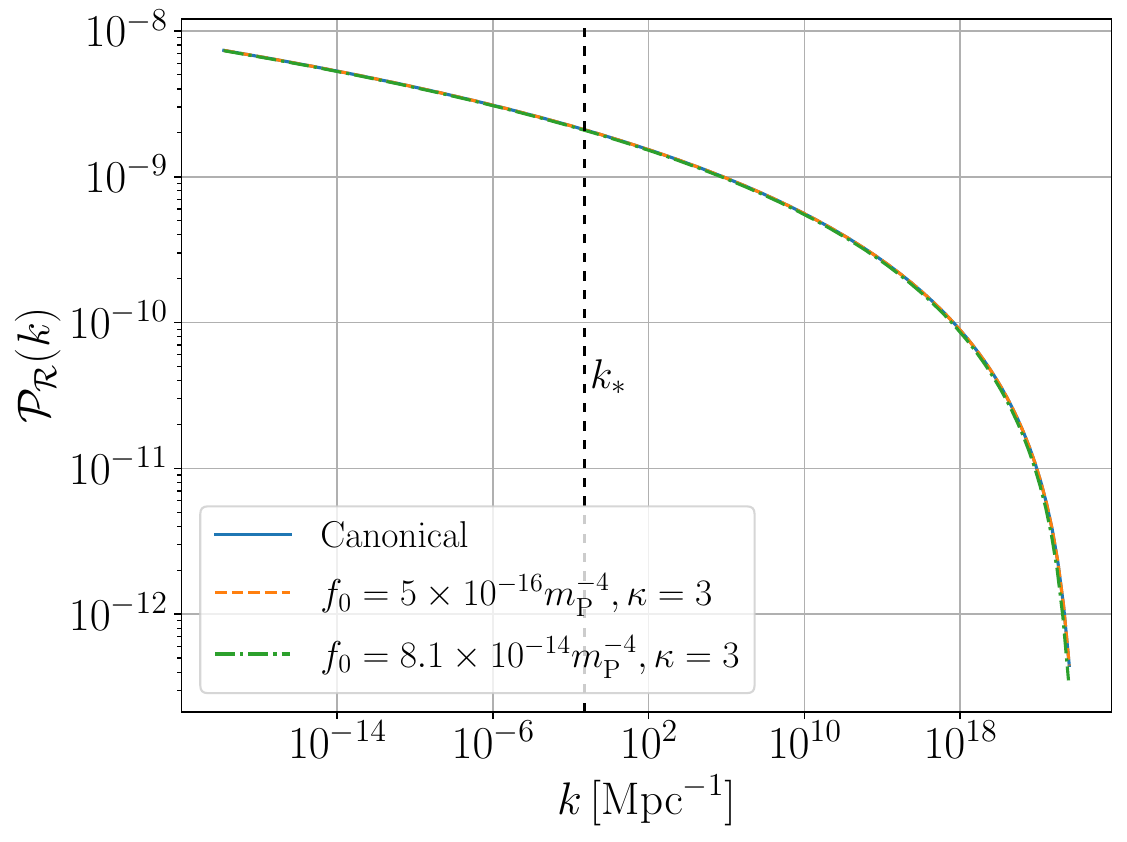}
     \end{subfigure}
     \begin{subfigure}[b]{0.4\textwidth}
         \centering
         \includegraphics[width=\textwidth]{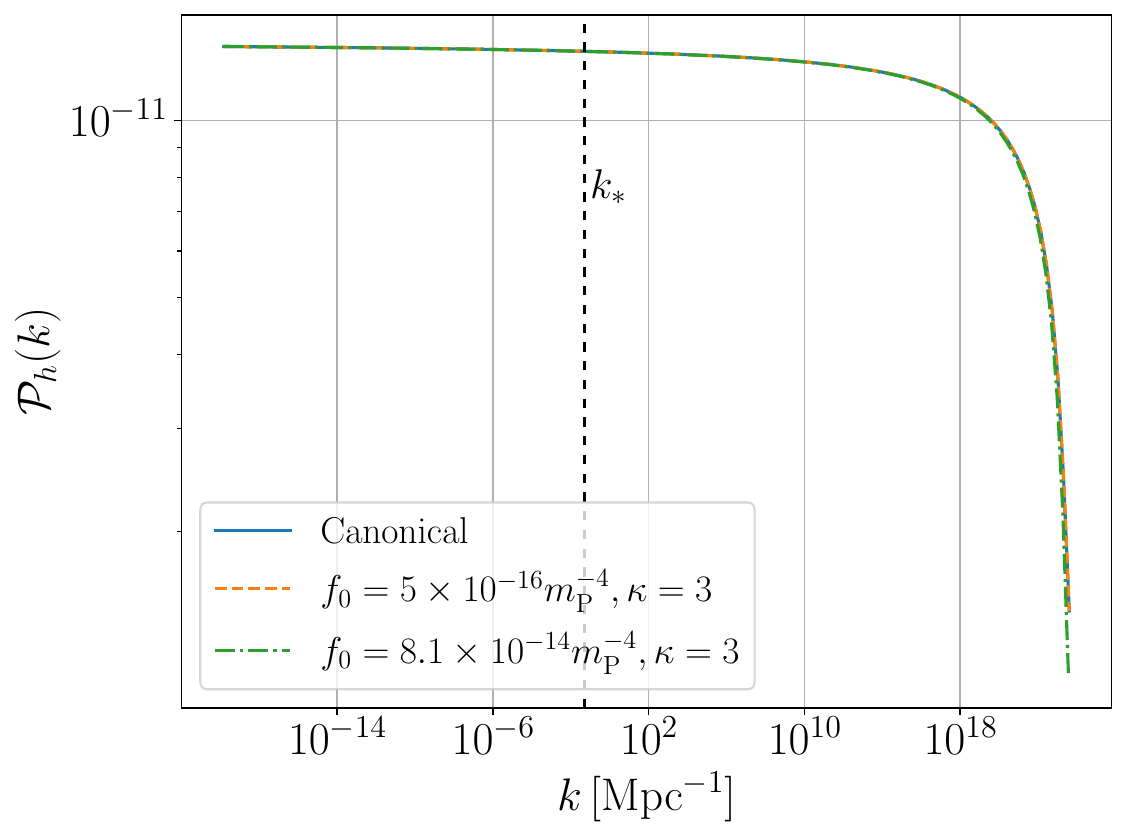}
     \end{subfigure}
     \caption{Scalar (top) and tensor (bottom) power spectra as a function of the wavenumber $k$ (in Mpc$^{-1}$). In blue the dynamics is governed by a canonical action, while in dashed it is governed by a DBI action with an exponential warp factor~\eqref{eq:f}, with $\{f_0\m^4,\kappa\}=\{5\times 10^{-16}, 3\}$ (orange) and $\{f_0\m^4,\kappa\}=\{8.1\times 10^{-14}, 3\}$ (green).}
     \label{fig:powerspectra}
\end{figure}

The top panel of Fig.~\ref{fig:powerspectra} shows the scalar primordial power spectra for the same parameter values $\{f_0,\kappa\}$ as in the discussion around Fig.~\ref{fig:hubbleFlow}. Again, small enough $f_0$ (orange curves) leads to dynamics that closely follow that of the canonical case (blue curves), while larger $f_0$ (green curves) leads to small differences at large $k$, for modes that leave the horizon near the end of inflation, when deviations of $c_{\rm s}$ from unity become larger. Differences between tensor primordial power spectra (bottom panel of Fig.~\ref{fig:powerspectra}) also remain small, consistently with the fact that their evolution is only affected by the DBI modifications at the background level.

\subsection{Post-inflationary evolution}
After the end of inflation, we solve numerically the KG equation~\eqref{eq:KG-DBI}, with the Friedmann equation containing the radiation component~\eqref{eq:1stFriedmannpostinfl} acting as a constraint. Initial conditions for the field are simply given by the end values of the numerical integration during inflation, while those of radiation are given by Eq.~\eqref{eq:rhore} below. We integrate until the system is deep in the radiation-dominated era, when $\rho_\ra = 10^7 \rho^{(\phi)}$. Throughout the early Universe evolution considered here, the pressureless matter component is negligible.

 \begin{figure}[ht] 
     \centering
     \begin{subfigure}[b]{0.4\textwidth}
         \centering
         \includegraphics[width=\textwidth]{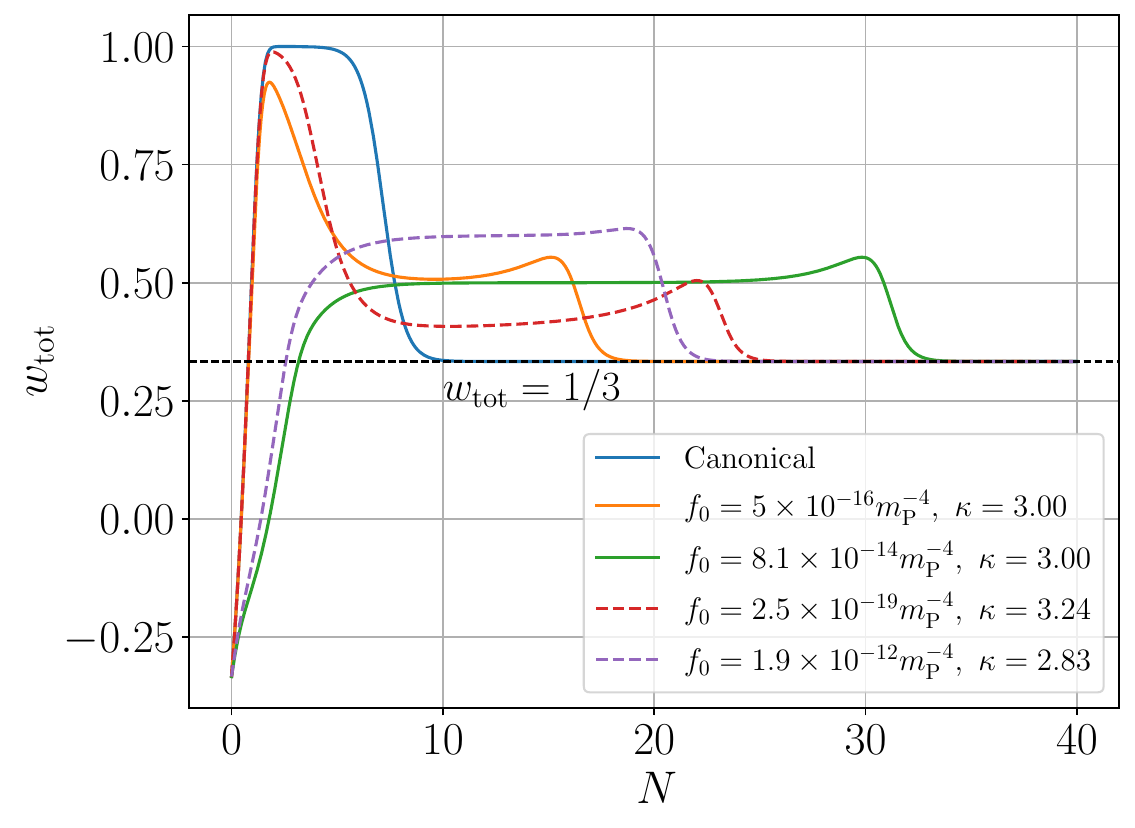}
     \end{subfigure}
          \centering
     \begin{subfigure}[b]{0.4\textwidth}
         \centering
         \includegraphics[width=\textwidth]{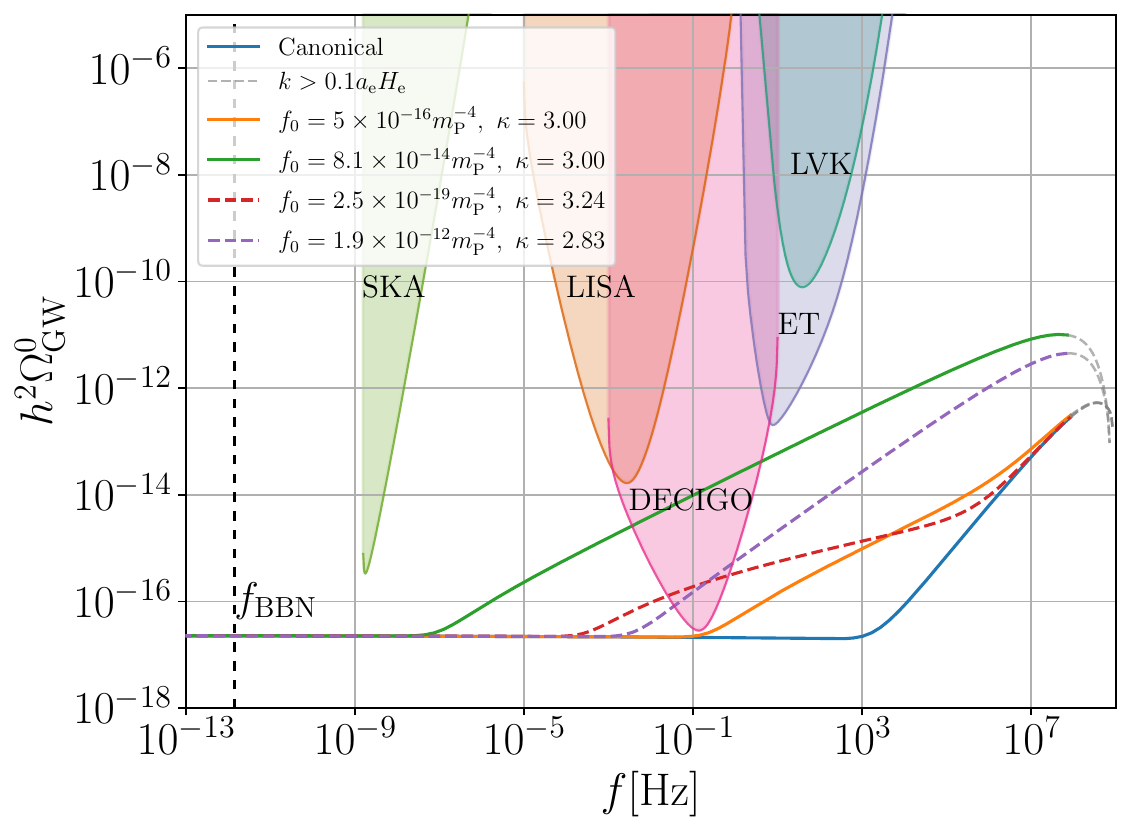}
     \end{subfigure}
     \caption{Top: Total EoS parameter $w_{\rm tot}$ as a function of the number of e-folds for different combinations of $\{f_0,\kappa\}$. The horizontal dashed line indicates $w_{\rm tot}=1/3$. Bottom: Corresponding density parameter of GWs as a function of frequency, superimposed with the PLICs of different GW experiments. The gray dashed curves indicate the UV portion of the spectra beyond the cutoff $k>0.1a_{\rm e}H_{\rm e}$.}
     \label{fig:w}
\end{figure}
In the top panel of Fig.~\ref{fig:w}, we show the total post-inflationary EoS parameter $w_{\rm tot}$~\eqref{eq:wtot} for different combinations of $\{f_0,\kappa\}$. In contrast to the canonical case (blue curve), for which $w_{\rm tot} \simeq 1$ during kination, the DBI models exhibit a stiff phase with $1/3<w_{\rm tot}<1$. During scalar-kinetic domination, the approximately constant value of $w_{\rm tot}$ agrees with the analytical estimate in Eq.~\eqref{eq:kappaintermsofw}. For fixed $\kappa$ (orange and green), the system spends more (less) time in the constant $w_{\rm tot}$ regime for larger (smaller) values of $f_0$. Varying $\kappa$ (dashed purple and red) changes the value of $w_{\rm tot}$ during the stiff phase.

In the bottom panel of Fig.~\ref{fig:w}, we show the present-day GW spectral energy density parameter corresponding to the background evolution in the top panel,\footnote{Since Eq.~\eqref{eq:rho_h_k_today} relies on the assumption that tensor modes have reached their constant super-horizon amplitude, we represent the part of the spectrum above the cutoff $k > 0.1 a_{\rm e}H_{\rm e}$ in dashed gray, emphasizing that predictions in this region are expected to be less accurate. Nevertheless, these uncertainties do not significantly affect our detectability conclusions.} obtained via Eq.~\eqref{eq:rho_h_k_today}. In the same figure, we also show the power-law integrated curves (PLICs)~\cite{Thrane:2013oya} for LVK~\cite{LIGOScientific:2019lzm}, DECIGO~\cite{Kawamura:2006up,Kawamura:2011zz,Kawamura:2020pcg}, ET~\cite{Punturo:2010zz,Hild:2010id}, LISA~\cite{Bartolo:2016ami,Caprini:2019pxz,LISACosmologyWorkingGroup:2022jok}, and SKA~\cite{Janssen:2014dka}. We find that the DBI non-canonical contributions indeed soften the blue tilt of canonical kination (blue line), as illustrated by all DBI curves. Furthermore, for the same value of $\kappa$, sufficiently long periods of constant-$w_{\rm tot}$ can turn an unobservable spectrum (orange line) into an observable one (green line). Different values of $\kappa$ and $f_0$ lead to characteristic spectral shapes (dashed red and purple lines). If detected, they may shed light on microphysics of post-inflationary stiff phases. 

 \begin{figure*}[ht] 
     \centering
     \includegraphics[width=0.7\textwidth]{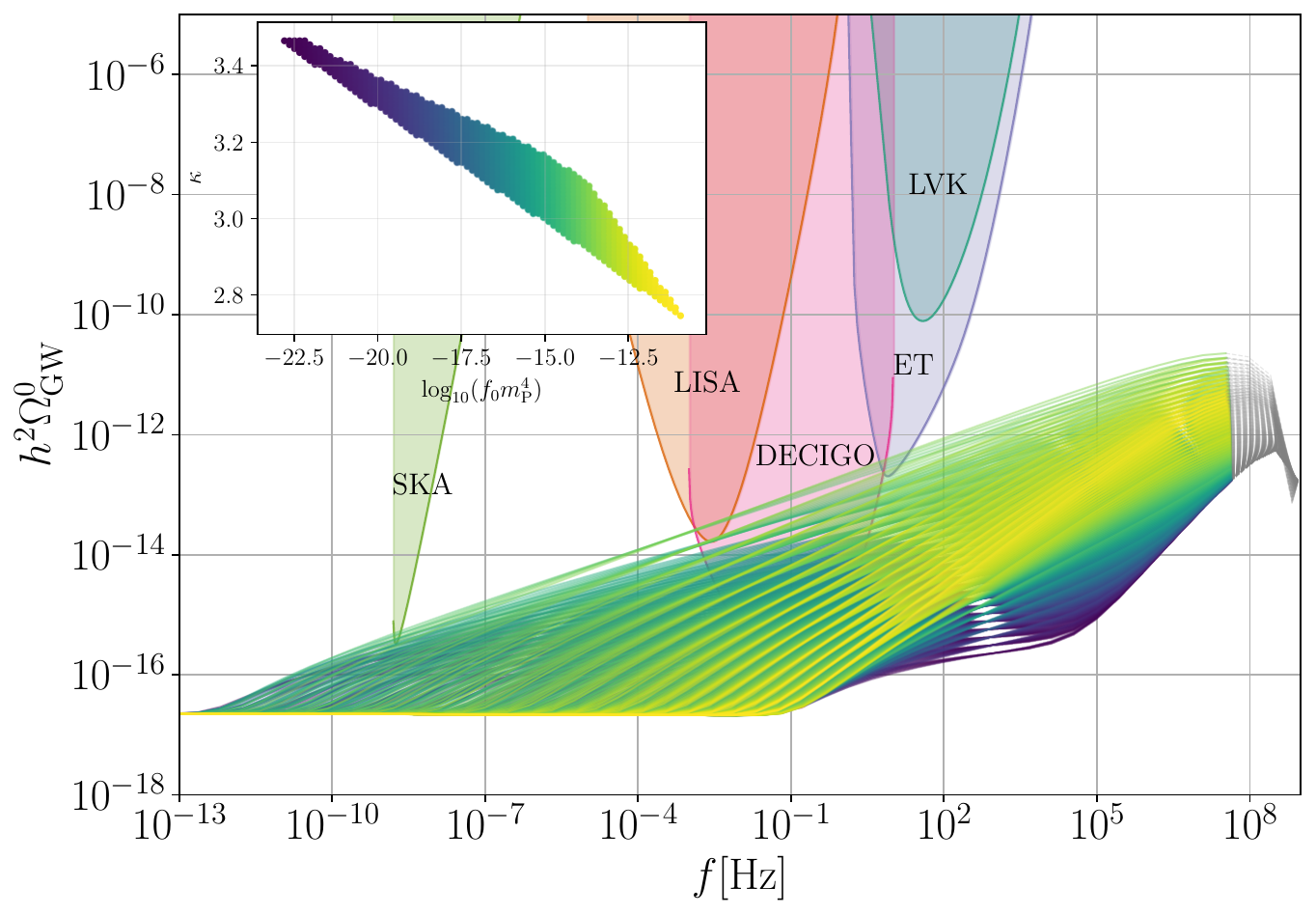}
     \caption{Surviving points in the $(\log_{10}(f_0\m^4), \kappa)$ plane together with the corresponding GW density spectra $h^2\Omega_{\rm GW}^0(f)$. All points satisfy the CMB constraints at the pivot scale~\eqref{eq:condition1}-\eqref{eq:condition4}, the BBN bound~\eqref{eq:BBNbound}, the reheating condition $\rho_{\rm r}(N_{\rm reh})=\rho^{(\phi)}(N_{\rm reh})$ taking place above $T_{\rm BBN}=0.1$MeV, together with $h^2\Omega_{\rm GW}^0(f)$ crossing at least one sensitivity curve of the GW experiments. The gray dashed curves indicate the UV portion of the spectra beyond the cutoff $k>0.1a_{\rm e}H_{\rm e}$.}
     \label{fig:scan}
\end{figure*}

\subsection{Parameter Space}
To assess the viability and observational prospects of our proposed mechanism, we explore the parameter space of the model. Our main goal is to investigate the effect of the DBI kinetic structure on the post-inflationary evolution and primordial GW spectrum. To this end, our general strategy is to fix $\{V_0,\lambda,n\}$ such that the canonical model $P(X,\phi)=X-V(\phi)$ yields a curvature power spectrum compatible with CMB observations. Then, keeping $\{V_0,\lambda,n\}$ fixed, we consider warp factor parameters $\{f_0,\kappa\}$ such that the DBI inflationary predictions around the CMB pivot scale remain within the Planck 2018 1$\sigma$ confidence limits (obtained when the running of the scalar spectral index is allowed to vary)~\cite{Planck:2018vyg}\footnote{Recent high-$l$ CMB measurements by the South Pole Telescope (SPT)~\cite{SPT-3G:2025bzu} and the Atacama Cosmology Telescope (ACT)~\cite{AtacamaCosmologyTelescope:2025blo}, when combined with BICEP/Keck~\cite{BICEP:2021xfz} and Planck 2018~\cite{Planck:2018vyg} yield a value for the scalar spectral index given by $n_{\rm s}=0.9682\pm 0.0032$~\cite{Balkenhol:2025wms} (68\%C.L.), slightly higher than the Planck 2018 result $n_{\rm s}=0.9649 \pm 0.0042$ (68\%C.L.). All three measurements remain statistically compatible when analyzed separately, although the combined posteriors show some dependence on likelihood choices and priors~\cite{Balkenhol:2025wms}. Since our goal is to isolate the effect of the DBI dynamics on the post-inflationary GW spectrum rather than perform a global fit of the inflationary parameters, we choose Planck 2018 as our reference.} and the 95\% C.L upper bound on the tensor-to-scalar ratio from BICEP/Keck~\cite{BICEP:2021xfz}
\begin{align}
    &n_{\rm s} = 0.9641\pm0.0044,\label{eq:condition1}\\
    &\log(10^{10}A_{\rm s})=3.047\pm0.015,\\
    &\alpha_{\rm s}=-0.0045\pm 0.0067,\\
    & r < 0.036\label{eq:condition4}.
\end{align}
We do not require the curvature power spectrum to remain unchanged at higher values of $k$, allowing the DBI contributions to modify the final stages of inflation as well as the subsequent post-inflationary evolution.

As a benchmark set of inflationary parameters, we take
\begin{equation}
    \label{eq:parametervalues}
    V_0 = 2.01443\times 10^{-10}\m^4,\quad \lambda=3\times 10^{-10},\quad n=8,
\end{equation}
together with
\begin{equation}
    \label{eq:rhore}
    \rho_\ra^{\rm e} = 8.1\times 10^{-17}\m^4.
\end{equation}
The above value for the radiation energy density at the end of inflation is a phenomenological choice based on previous quintessential inflation studies with successful reheating~\cite{Ahmad:2019jbm}. It fixes the scale factor at the end of inflation, and therefore the time at which the pivot scale $k_{*}=0.05$Mpc$^{-1}$ exits the horizon during inflation, corresponding roughly to $N_{*}=-59.4$ and $\phi(N_{*})=10.2\m$ for the representative models considered in this section. 

With the above values, we obtain results in the canonical case compatible with observations (see Fig.~\ref{fig:powerspectra})
\begin{align}
    &A_{\rm s}(k_{*}) = 2.10\times10^{-9},  &n_{\rm s}(k_{*}) = 0.9612,\nonumber\\
    &\alpha_{\rm s}(k_{*}) = -6.48\times10^{-4}, &r=0.0062.
\end{align}

There are additional constraints the model must satisfy. First, we require that the Universe is radiation-dominated by the time BBN takes place. In other words, $\rho_\ra(N_{\rm reh})>\rho_{\rm BBN}=\pi^2 g_{*}(T_{\rm BBN})T_{\rm BBN}^4/30 \simeq 3.14\times10^{-90}\m^4$, where we have used $T_{\rm BBN}\simeq 0.1\text{MeV}$~\cite{Caprini:2018mtu}. Furthermore, the produced GW density $\langle\hat\rho_h\rangle$ must not exceed the limits on the abundance of radiation during BBN. This bound is usually expressed in terms of extra neutrino species $\Delta N_{\rm eff}$ as~\cite{Caprini:2018mtu,Aggarwal:2025noe}
\begin{equation}
    \label{eq:BBNbound}
    \int_{f_{\rm BBN}}^{f_{\rm e}} \frac{\text{d}f}{f}h^2\Omega_{\rm GW}^0(f)<5.6\times 10^{-6}\Delta N_{\rm eff}< 6.0\times10^{-7},
\end{equation}
where integration takes place for modes that were already sub-horizon at the time of BBN, and we have taken $\Delta N_{\rm eff}<0.107$ (95\% C.L.)~\cite{Goldstein:2026iuu}. Using $T_{\rm BBN}\simeq 0.1\text{MeV}$, the present-day frequency corresponding to the comoving Hubble scale at BBN is $f_{\rm BBN}=1.42\times 10^{-12}$Hz. Lastly, since LVK has not detected a stochastic background of GWs, we also discard $\{f_0,\kappa\}$ values that lead to a signal within its sensitivity band. For consistency, we additionally impose that the background does not undergo a second period of accelerated expansion following inflation.\footnote{For some extreme parameter values, the non-canonical kinetic contributions can be so strong that the background evolution resumes acceleration at some point after the end of inflation.}

For our parameter scan, in addition to the above constraints, we also require that the resulting GW spectra cross the sensitivity curve of at least one of the GW experiments. As for the scan itself, we compute $\Omega_{\rm GW}^0(f)$ for combinations of $\kappa$ and $f_0$, with $\kappa$ taking 100 evenly spaced values within the interval $\sqrt{12} > \kappa > \sqrt{6}$, and $\log_{10} (f_0 \m^4)$ taking 100 evenly spaced values within the logarithmic interval $-9 > \log_{10} (f_0 \m^4) > -24$. The boundaries of the interval for $\kappa$ are motivated from the requirement that the post-inflationary period has a stiff phase, see Eq.~\eqref{eq:kappa_limits}. Those for $\log_{10} (f_0 \m^4)$ are chosen such that the entire surviving parameter region is contained within them (we have also run scans with broader limits, showing no difference). 

In Fig.~\ref{fig:scan} we show the 835 surviving points satisfying all conditions mentioned above, in the $(\log_{10}(f_0\m^4), \kappa)$ plane, together with the associated GW density spectra. There exists ample parameter space, with $f_0\m^4$ spanning 10 orders of magnitude for the chosen range in $\kappa$ leading to a stiff phase. Preliminary tests show this trend is maintained for values of the inflationary parameters different from Eqs.~\eqref{eq:parametervalues}-\eqref{eq:rhore}, although a parameter scan over the parameter space of the full model is beyond the scope of the present work.

\section{Conclusions} \label{sec:conclusions}
In this work, we studied the effect on the spectral density of GWs, $\Omega_{\rm GW}^0(f)$, of non-canonical kinetic terms in the action of a scalar field that determines the post-inflationary background evolution. To do so, we considered a setup combining NO inflation with a DBI action. We showed the non-canonical terms contribute to the dynamics in such a way that the EoS parameter of the Universe during post-inflationary stiff phases is reduced relative to the canonical case, leading to a slope for $\Omega_{\rm GW}^0(f)$ that is in general smaller.

In particular, we showed that the EoS parameter $w$ of the Universe during a period dominated by the energy density of a kinetically dominated scalar field governed by a DBI action is $w<1$, generally smaller than the canonical kination result $w=1$. The amount by which $w$ deviates from unity, as well as the length of the period, depend on the warp factor of the brane $f(\phi)$ and on $X$. In order to test our claims, we chose an exponential function $f(\phi)=f_0 e^{\kappa \phi/\m}$, capable of producing a period with constant EoS parameter for certain combinations of parameter values. 

The slope of $\Omega_{\rm GW}^0(f)$ depends on the background dynamics in a non-trivial way via $w$, taking values that range from 0 to 1 for values of $w$ between $1/3$ and $1$ (neglecting the small primordial tensor tilt). Therefore, the characteristic blue tilt in the GW spectrum, for tensor modes that re-enter the horizon during canonical kination, is in general reduced by the non-canonical contributions of the DBI action. We find two competing effects. First, the smaller value of $w$ reduces the blue tilt, and thus the enhancement, of GW modes re-entering the horizon during the stiff phase. Second, since the density of the scalar field now redshifts at a slower rate than in the canonical case, the onset of radiation domination is delayed, prolonging the stiff period and extending the range of frequencies in $\Omega_{\rm GW}^0(f)$ affected by it.  The interplay between these two effects has observational consequences, allowing GW spectra that would otherwise be inaccessible by future GW experiments to become potentially detectable. It also highlights the possibility of testing the kinetic structure of the scalar-field Lagrangian with GW measurements. 

We tested these predictions by numerically solving the inflationary background and perturbation equations, followed by the post-inflationary background evolution, and computing the resulting present-day GW spectra. The numerical results agree with the analytical expectations, showing that the EoS parameter during the stiff phase is primarily controlled by $\kappa$, while its duration can be varied through $f_0$.

We also studied the robustness of the mechanism by carrying out a parameter scan of the DBI parameters. To do so, we fixed the parameters of the potential as well as the density of radiation at the end of inflation, without specifying the reheating mechanism that produces it. These parameters are fixed such that the canonical model $P=X-V$ predicts an amplitude, tilt, and running of the scalar power spectrum, at the CMB pivot scale, compatible with the Planck 2018~\cite{Planck:2018vyg} 1$\sigma$ confidence limits, together with a tensor-to-scalar ratio compatible with the BICEP/Keck~\cite{BICEP:2021xfz} 95\% C.L. upper bound. Then, we solved the DBI model for different combinations of $\kappa$ and $f_0$, keeping only those that retain the correct CMB predictions and achieve radiation domination before BBN, in combination with GW density spectra that satisfy the BBN bound and that cross the sensitivity curves of at least one of the future GW experiments. The results show that for values of $\kappa$ leading to a post-inflationary stiff phase, $f_0$ can take values spanning around 10 orders of magnitude, demonstrating that the mechanism remains viable over a substantial region of the explored parameter space.

Our results therefore highlight the potential of primordial GW observations to probe the kinetic structure of the scalar field governing the post-inflationary evolution, even when different theories lead to nearly indistinguishable predictions at CMB scales.

Looking ahead, it would be interesting to perform a Bayesian analysis of the full model, also including the inflationary parameters, by combining different datasets including Planck 2018, ACT, and SPT, together with forecasts for future GW experiments. Furthermore, since we find departures of $c_{\rm s}$ from unity at small scales for parameter combinations leading to potentially observable GW signals, it would be interesting to investigate the corresponding primordial non-Gaussianities and their possible implications for primordial black hole formation and scalar-induced gravitational waves, providing additional observational probes for the model.

\appendix
\section{DBI stiff background with evolving sound speed} \label{Appendix}

In this section we investigate analytically the post-inflationary dynamics of the DBI field, focusing on its ability to support a sufficiently long stiff phase. Owing to the complicated nonlinear dynamics of DBI models, to make analytical progress we will follow a limited approach based on the constant-roll (CR) condition and a suitable ansatz for $\gamma$. 

\begin{figure}[H] 
    \centering
    \includegraphics[width=0.7\columnwidth]{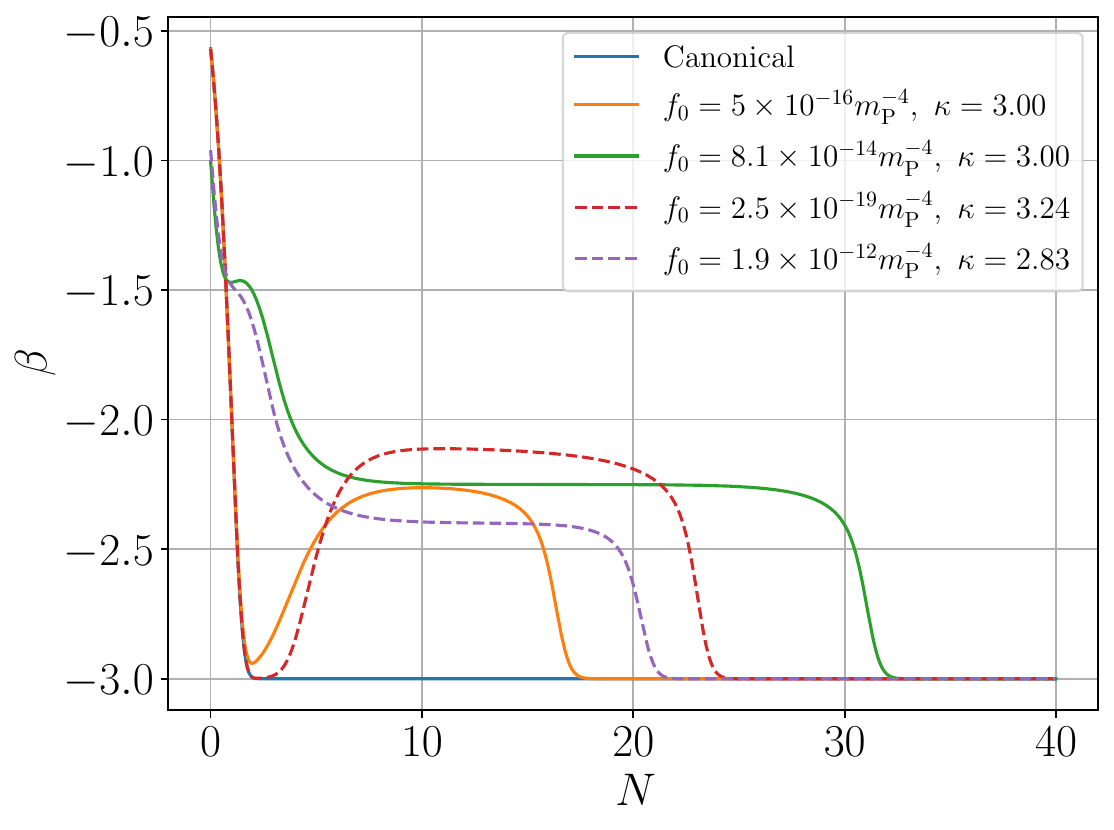}
     \caption{Post-inflationary evolution of $\beta$ for the same parameter values $\{f_0,\kappa\}$ as in Fig.~\ref{fig:w}.}
     \label{fig:CR}
\end{figure}

In Sec.~\ref{sec:postinflation}, our discussion revolved around the kinetically dominated regime with constant $w^{(\phi)}$, and hence constant $c_{\rm s}$.\footnote{For treatments of constant speed of sound and power law backgrounds during inflation see Refs.~\cite{Spalinski:2007un,Spalinski:2007dv}.} Interestingly, it corresponds to a CR evolution. Indeed, defining the CR parameter
\begin{equation}
    \frac{\ddot{\phi}}{H\dot{\phi}} \equiv \beta,
\end{equation}
and using the Friedmann equations in the kinetically dominated regime, together with the constancy of $\gamma$, we obtain
\begin{equation}
    \beta \overset{X\gg V}{\simeq} \frac{\dot{H}}{H^2} = -\frac{3}{2}\left(1+\frac{1}{\gamma}\right) = - \frac{3}{2}\frac{\kappa^2}{\kappa^2 - 3}.
\end{equation}
The validity of this approximation is confirmed by the numerical evolution shown in Fig.~\ref{fig:CR}, where $\beta$ approaches the analytical prediction during the stiff period. Motivated by this, we investigate a more general class of CR DBI backgrounds in which the sound speed is allowed to evolve.

\subsection{Constant-roll dynamics}
Taking a derivative of $\epsilon_1=-\dot{H}/H^2$, using the definition of the Hubble flow parameters~\eqref{eq:flowparams}, and the Raychaudhuri equation, it is straightforward to obtain
\begin{equation}
    \label{eq:betaCR}
    \beta = \frac{1}{2}(\epsilon_2 - \delta_1) - \epsilon_1.
\end{equation}
This relation is exact in the absence of matter components other than the scalar field, without relying on kinetic domination or constant speed of sound. It can be expressed in a more convenient way by using
\begin{equation}
    \label{eq:eps1CR}
    \epsilon_1 = \frac{2\m^2}{\gamma}\left(\frac{H_\phi}{H}\right)^2,
\end{equation}
\begin{equation}
    \epsilon_2 = 2\epsilon_1 - \frac{4\m^2}{\gamma}\frac{H_{\phi\phi}}{H} + \frac{2\m^2}{\gamma}\frac{H_\phi}{H}\frac{\gamma_\phi}{\gamma},
\end{equation}
and
\begin{equation}
    \delta_1 = \frac{\dot{\gamma}}{H\gamma} = -\frac{2\m^2}{\gamma}\frac{H_\phi}{H}\frac{\gamma_\phi}{\gamma},
\end{equation}
where we have used Eqs.~\eqref{eq:2ndHamJac} and~\eqref{eq:flowparams}. Putting everything together, Eq.~\eqref{eq:betaCR} can be written as
\begin{equation}
    H_{\phi\phi}-\frac{\gamma_\phi}{\gamma}H_\phi+\frac{\gamma\beta}{2\m^2}H=0.
\end{equation}

\subsubsection{Exponential growth for $\gamma$}
To investigate the effect of an evolving sound speed while retaining analytical tractability, we impose the CR condition $\beta=\mathrm{const.}$ and we adopt the exponential ansatz
\begin{equation}
    \gamma(\phi) = \gamma_0 e^{\alpha\phi/\m},
\end{equation}
where $\gamma_0$ and $\alpha$ are positive constants. For $\beta>0$, it leads to the solution
\begin{equation}
    H(x) = c_1 x J_1(x) + c_2 x Y_1(x), \quad x \equiv \frac{\sqrt{2\beta\gamma}}{\alpha},
\end{equation}
where $c_1$ and $c_2$ are integration constants and $J_1$ and $Y_1$ are Bessel functions of the first and second kind, respectively. For $\beta<0$, we instead find
\begin{equation}
    \label{eq:analyticalsol}
    H(y) = c_1 y I_1(y) + c_2 y K_1(y),\quad y \equiv \frac{\sqrt{2\abs{\beta}\gamma}}{\alpha},
\end{equation}
where $I_1$ and $K_1$ are modified Bessel functions of the first and second kind, respectively. 

We examine the behaviour of the solution with $\beta<0$, which, as we show below, can sustain a prolonged stiff phase. Using Eq.~\eqref{eq:2ndHamJac}, it is easy to see that the first mode in Eq.~\eqref{eq:analyticalsol} corresponds to $\dot{\phi}<0$, while the second one corresponds to $\dot{\phi}>0$. For simplicity, we set $c_1=0$ consistently with our field evolution convention and write
\begin{equation}
    \label{eq:Hcr}
    H(y)=\frac{H_{\rm cr}}{y_{\rm cr}K_1(y_{\rm cr})}\,yK_1(y),
\end{equation}
where the subscript `cr' indicates the time when the scalar field engages in the constant-roll regime after the end of inflation. The EoS parameter for this background immediately follows as
\begin{equation}
    w = \frac{2}{3}\epsilon_1 - 1 = \frac{2\abs{\beta}}{3}\left(\frac{H_y}{H}\right)^2 - 1 = \frac{2\abs{\beta}}{3}\frac{K_0^2(y)}{K_1^2(y)} - 1,
\end{equation}
where we have used Eq.~\eqref{eq:eps1CR} and $\text{d}y/\text{d}\phi = \alpha y /(2\m)$. Using the asymptotic limits of the modified Bessel functions, it follows that $w\overset{y\ll1}{\simeq}-1$ and  $w\overset{y\gg1}{\simeq}2\abs{\beta}/3 - 1$. Therefore, given $2<\abs{\beta}\leq 3$, this background formally interpolates between an inflationary solution and a stiff phase with $1/3<w\leq 1$. However, the physically allowed solution is restricted by further requirements, as we now discuss.

Using Eq.~\eqref{eq:1stHamJac}, the potential may be expressed as
\begin{equation}
    \label{eq:potentialApp}
    V = 3\m^2 H^2\left[1-\frac{2\epsilon_1}{3}\frac{\gamma}{1+\gamma}\right].
\end{equation}
For $2<\abs{\beta}\leq 3$, combining the stiff phase requirement $\epsilon_1>2$ with the positivity of the potential $\epsilon_1<3(\gamma+1)/(2\gamma)$, we find the allowed range of $y$
\begin{equation}
    \label{eq:anotherbound}
    2 < \abs{\beta} \frac{K_0^2(y)}{K_1^2(y)} < \frac{3}{2}\left(1+\frac{1}{\gamma(y)}\right),
\end{equation}
where $\gamma(y)=\alpha^2y^2/(2\abs{\beta})$. Furthermore, since $\gamma\geq1$, we have the additional restriction
\begin{equation}
    \label{eq:bound}
    y\geq y_{\rm min} \equiv \frac{\sqrt{2\abs{\beta}}}{\alpha}.
\end{equation}

Finally, to study whether the model allows for a sustained stiff phase, we compute the integrated expansion during the constant-roll regime. Using Eqs.~\eqref{eq:2ndHamJac} and~\eqref{eq:Hcr} we obtain
\begin{align}
    N_{\rm max}&=\frac1{|\beta|}\int_{y_{\rm cr}}^{y_{\rm max}}\text{d}y\frac{H}{|H_y|}=\frac1{|\beta|}\int_{y_{\rm cr}}^{y_{\rm max}}\text{d}y\frac{K_1(y)}{K_0(y)}=\nonumber\\
    &=\frac{1}{\abs{\beta}}\log\left[\frac{K_0(y_{\rm cr})}{K_0(y_{\rm max})}\right],
\end{align}
where we have used $K_0'(y)=-K_1(y)$ in the last step. Taking the representative value $\abs{\beta}=2.3$, we calculate the maximum expansion by setting $y_{\rm cr}=y_{\rm min}$, provided the stiff-phase condition is satisfied at this boundary, and taking $y_{\rm max}$ at the limiting value where $V(y_{\rm max})\to 0^+$. For $0.017\leq\alpha\leq0.079$, we obtain a maximum expansion within $5\leq N_{\rm max}\leq20$. The corresponding expansion-averaged EoS
\begin{align}
\bar{w}&=\frac{1}{N_{\rm max}}\int_{y_{\rm cr}}^{y_{\rm max}} \text{d}y\frac{dN}{dy} w(y)  =\nonumber\\
&=-1 + \frac{2}{3N_{\rm max}}\int_{y_{\rm cr}}^{y_{\rm max}}\text{d}y\frac{K_0(y)}{K_1(y)}=\nonumber\\
&=-1 + \frac{2}{3 N_{\rm max}}\log\left[\frac{y_{\rm cr} K_1(y_{\rm cr})}{y_{\rm max} K_1(y_{\rm max})}\right],
\end{align}
where we have used $(y K_1(y))'=-yK_0(y)$ in the last step, lies in the range $0.487\leq\bar{w}\leq0.523$.

\newpage

\section*{Acknowledgements}
SSL would like to thank the Indo-French Centre for the Promotion of Advanced Research (IFCPAR/CEFIPRA) for support of the proposal 6704-4 titled `Testing flavors of the early universe beyond vanilla models with cosmological observations’ under the Collaborative Scientific Research Programme. ET is supported by the ``Fonds de
la Recherche Scientifique'' (FNRS) under the IISN grant
number 4.4517.08.

\bibliography{bibliography}

\end{document}